\documentclass{aa}  
\usepackage{graphicx}
\usepackage{natbib}
\usepackage{txfonts}
\begin{document}
   \title{Building merger trees from cosmological $N$-body simulations}

   \subtitle{towards improving galaxy formation models using subhaloes}

   \author{D. Tweed\inst{1}\inst{2}
          \and
          J.Devriendt\inst{1}\inst{2}
          \and
          J. Blaizot\inst{1}
          \and
          S. Colombi\inst{3}
          \and
          A. Slyz \inst{2}
        }
 
%   \offprints{}

        \institute{Universit\'e de Lyon, Lyon, F-69003, France; \\
          Universit\'e Lyon~1, Villeurbanne; \\
          CNRS, UMR 5574, Centre de Recherche Astrophysique de Lyon; \\
          Observatoire de Lyon, Saint-Genis Laval, F-69230; \\
          Ecole Normale Sup\'erieure de Lyon, F-69007
          \and
          Astrophysics, University of Oxford, Keble Road, Oxford OX1 3RH, UK; 
          \and
          Institut Astrophysique de Paris (IAP)\\
          98 bis Boulevard d'Arago, F-75014 Paris, France
          }
 
   \date{Received month day, year; accepted month day, year}

% \abstract{}{}{}{}{} 
% 5 {} token are mandatory
 
  \abstract
   {In the past decade or so, using numerical $N$-body simulations to describe the
     gravitational clustering of dark matter (DM) in an expanding universe 
     has become the tool of choice for tackling the issue of hierarchical galaxy formation. 
     As mass resolution increases with the power of supercomputers, one is able 
     to grasp finer and finer details of this process, resolving more and more
     of the inner structure of collapsed objects. This begs one to revisit
     time and again the post-processing tools with which one transforms particles into
     ``invisible'' dark matter haloes and from thereon into luminous galaxies.}
   {Although a fair amount of work has been devoted to growing Monte-Carlo merger trees
     that resemble those built from an $N$-body simulation, comparatively
     little effort has been invested in quantifying the caveats one necessarily encounters 
     when one extracts trees directly from such a simulation. To somewhat revert the tide, 
     this paper seeks to provide its reader
     with a comprehensive study of the problems one faces when following this route.}
   {The first step in building merger histories of dark matter haloes and their subhaloes 
     is to identify these structures in each of the time outputs (snapshots) produced by 
     the simulation. Even though we discuss a particular implementation of such
     an algorithm (called AdaptaHOP) in this paper, we believe that our results do not depend on the 
     exact details of the implementation but instead extend to most if not all (sub)structure 
     finders. 
     To illustrate this point in the appendix we compare AdaptaHOP's results to the standard friend-of-friend (FOF) algorithm, widely utilised in the astrophysical community. We then  highlight different ways of building merger histories from AdaptaHOP haloes and subhaloes, contrasting their various advantages and drawbacks.}
   {We find that the best approach to (sub)halo merging histories 
     is through an analysis that goes back and forth between identification 
     and tree building rather than one that conducts a straightforward sequential
     treatment of these two steps. This is rooted in the complexity of the merging trees
     that have to depict an inherently dynamical process from the partial 
     temporal information contained in the collection of instantaneous
     snapshots available from the $N$-body simulation. However, we also propose 
     a simpler sequential ``Most massive Substructure Method'' (MSM)
     whose trees approximate those obtained via the more complicated non sequential method.}
    {}


   \maketitle
%__________________________________________________________________

\section{Introduction}

Dark matter haloes and their mass assembly histories are the fundamental 
bricks of any non linear structure formation theory based on the current concordance ($\Lambda$CDM) 
model which has been so successful at reproducing large scale structure data
\citep{Dunkley2009}. It is therefore natural that a lot of effort 
has been devoted to finding semi-analytic descriptions of this process.
These culminated with the seminal papers on the extended Press Schechter (EPS) formalism 
\citep{Bond1991,LaceyCole1994}, as it became possible to make Monte-Carlo 
realisations of merging histories of haloes using EPS.
However, it was also realised early on that shortcomings of the 
EPS theory needed to be circumvented (non spherical collapse, loss 
of internal structure, no spatial information) to get accurate halo mass distributions
and merging tree histories \citep{ShethLemson1999,Somerville1999}.

For a detailed critique of the EPS
theory, we refer the interested reader to \citet{Benson2005}, however we  
point out one of the most worrisome of its shortcomings. It may seem legitimate 
to generate merging trees for a representative sample of haloes
at a given redshift and then attempt to construct the halo mass function 
at an earlier time, combining the branches of these Monte-Carlo merging trees, with 
each branch appropriately weighted according to the EPS theory. However, in 
doing so, one will not obtain good agreement between this mass function and the mass function of haloes extracted directly 
from $N$-body simulations at this early epoch.
This discrepancy can be tuned   
empirically, but there is no theoretical justification as to why such 
a correction should be made \citep{Benson2001}.
This explains why people calibrate Monte-Carlo merging trees against those generated 
using $N$-body simulations, as done recently by \citet{NeisteinDekel2008a} and \citet{Parkinson2008}. 
Indeed merging trees directly built from $N$-body 
simulations naturally circumvent the shortcomings of the Monte-Carlo methods.
Moreover with the democratisation of (super)computer power
$N$-body simulations are becoming more and more available and resolved, and 
this implies that it will inevitably make more and more sense to build trees 
directly from them in the future.

However, as underlined in the hierarchical galaxy formation primer of \citet{Baugh2006},
the construction of a merger tree from the outputs of an $N$-body simulation
is not a trivial matter. The mass of a halo can decrease with time since haloes may 
spatially overlap one another at a given time output and therefore be blended 
together by the group-finding algorithm, then separate at the next time output for good, or come back together again later on. 
This paper is therefore devoted to identifying and quantifying the occurrences of these ``anomalies''
that plague $N$-body merger trees. It also proposes different methods of dealing with them
and contrasts/compares their advantages and disadvantages.

Its outline is as follows: in the first part (Sect. 2) we discuss the issue of dark 
matter halo and subhalo detection in cosmological $N$-body simulations; in the second (Sect 3) 
we build $N$-body merger histories based on three methods we use to construct subhaloes. We use these 
merger trees to pitch these methods against one another. We draw our conclusions in the third part 
(Sect. 4).

%__________________________________________________________________

\section{Dark matter halo and subhalo detection}

Most algorithms commonly used to identify dark matter haloes in $N$-body cosmological simulations 
are based either on a percolation algorithm, as the so called friend-of-friend (FOF)
algorithm \citep{Huchra1982,Davis1985} or a prescription to identify local 
maxima of the density field (e.g. DENMAX, \citet{Gelb1994}; SOD, \citet{LaceyCole1994}; 
BDM, \citet{KlypinGottloeber1999}; HOP, \citet{Eisenstein1998}).
With computational power rapidly increasing, these algorithms have been recently extended to 
probe the inner structure of the haloes and detect subhaloes within them
(e.g. IsoDen, \citet{Pfitzner1996}; SKID, \citet{Ghigna1998}; HFOF,
\citet{KlypinGottloeber1999}; SUBFIND, \citet{Springel2001}; AdaptaHOP,
\citet{Aubert2004}; MHF, \citet{Gill2004} and its successor AHF, \citet{Knollmann2009}).
 
Note however that, in the best of cases, these algorithms proceed in two consecutive steps: first they identify the 
halo in real (3 dimensional (3D)) space, and then they use velocity space information to ``refine'' 
the composition of their haloes (i.e. decide if a particle is gravitationally bound to it or not).
Ultimately, to obtain the most reliable results, one would want to define haloes as structures 
detected directly in the 6 dimensional (6D) phase space (for a review and extensive comparison of the methods which have been 
proposed to do that see \citet{Maciejewski2009a}), but the developments in that direction are pretty recent \citep{Diemand2006,Maciejewski2009b} so our approach in this paper remains three dimensional. Moreover, in practise, the bound structures detected in 6D space are not very different from the 3D ones, except that they tend to be 
systematically (albeit slightly) more massive \citep{Maciejewski2009b}.

In light of the previous comments, we feel it is a fair claim to say that none of the 
3D algorithms are completely satisfactory, and that the results of the analysis of 
any output of a cosmological $N$-body simulation in terms of halo/subhalo detection will, to a
certain extent, depend on the choice of algorithm used to
perform that detection. Bearing these limitations in mind we choose AdaptaHOP \citep{Aubert2004} 
as our halo and subhalo finder in this work. 

We first start by a brief description of this algorithm.
 We then discuss the advantages and 
disadvantages arising from three natural but different
methods to select subhaloes with AdaptaHOP: the Density Profile Method (DPM), the Most massive Sub-node Method (MSM) 
and the Branch History Method (BHM).   

\subsection{Dark matter halo detection}

In this section, we will be concerned with the way one can split 
an ensemble of particles in $N$-body simulation snapshots into DM haloes and
subhaloes. In other words, we want to group together particles on the
basis of the instantaneous values of their positions (and velocities) alone, 
using the a priori knowledge gleaned from the $N$-body simulation itself 
that positions contain the most accurate information. However, already at this
point, we emphasise that the merger history of haloes are imprinted in
the particle distribution. For instance, after a merger with another halo, a structure
can survive as a subhalo which will be present as a local density maxima within
the particle distribution of its host halo. 

In the appendix, we compare in details the most widely used halo finder FOF to the
AdaptaHOP algorithm that we will use to monitor substructure. Here, we simply point out 
that FOF consists in grouping together particles which
are closer than a distance $\epsilon = b \times $(mean inter-particle distance). Usually
$b$ is chosen to be $0.2$, which closely matches an average halo over-density of
$178 \times \rho_c$ (where $\rho_c$ is the critical density, i.e. the matter density necessary for the Universe to
be flat in the absence of other energy sources), 
obtained when solving the classic spherical collapse of a ``top-hat''
density perturbation in an Einstein-de Sitter universe. To optimise the mass resolution of the $N$-body
simulation,  haloes containing at least 20 particles are considered
as bound objects. Obviously, using such a threshold is not the best way to select
bound structures, because (i) it ignores the kinetic and potential energy of 
particles (ii) even if it was energetically justified, a halo sitting exactly 
on the threshold could still ``lose'' a particle by interacting with its environment and not be detected in the following
snapshot of the simulation. However, computing the potential energy of each
particle belonging to a halo is very expensive CPU wise, and at the same time
inaccurate due to both the pre-selection process of the halo members, which 
is necessarily based on a somewhat arbitrary spatial/velocity density 
cut, and the necessity to make do with a small number of particles per halo 
to maximise the halo mass range spanned by the simulation.
This experimental threshold of 20 is justified a posteriori by the fact that provided 
the time span between snapshots is not much larger than 200 Myrs, 
less than 30 \% of haloes with this many particles are lost
between two consecutive outputs (see e.g. Fig. \ref{small_halo_lost}). 

In principle, the simple FOF algorithm can be used to detect substructures, simply by running
with smaller and smaller values for the linking length \citep{KlypinGottloeber1999}. 
However, this method is quite inefficient because a
whole range of values for $b$ needs to be explored since real substructures (i) are
embedded within one another and (ii) have different density contrasts due to a
different epoch of collapse and tidal encounters. Furthermore, there exists no obvious
physical criterion to pick the ``best'' value of $b$ corresponding to a global 
level of substructure, as such a criterion would depend on merging history and therefore is
prone to vary from object to object.

Instead of relying directly on single particle positions to identify
substructure as the FOF would do, we can go a small step further and remark 
that substructures are going to be located at local maxima of their host
halo density field, so that the natural way to detect them is to compute this
density field. This simple assertion constitutes the core of the AdaptaHOP algorithm 
(as well as that of many others that we listed earlier) which can roughly be summarised by the
following steps (see Appendix
B of \citet{Aubert2004} for details) : 

\begin{enumerate}
\item For each particle in the $N$-body simulation, find the $n$ closest neighbours 
  using your favourite oct-tree algorithm ($n$ has a typical value comprised between 20 and 64, we use 20
  in this paper). The density $\rho_i$, associated to particle
  $i$ of mass $m_i$ is then computed using the following equation:
  \begin{equation}
    \rho_i = \frac{V_{box}}{V_r}\left( m_i + \sum_1^n m_j*spline\left( \frac{r_{ij}}{r} \right)\right)
    \label{sphdens}
  \end{equation}
  Here $m_j$ is the mass of particle $j$ (one of $i$'s $n$ closest
  neighbours), $r_{ij}$ is the distance between
  particles $i$ and $j$, and $r= 0.5 \times max(r_{ij}\mbox{, }j\in\{1,n\})$
  is the SPH smoothing length for particle $j$. $V_{box}$ and $V_r$ are the
  volumes of the simulation box and of a sphere of radius $r$ respectively, so
  their ratio yields the normalisation of the density across the box. The $spline$ function is the well-known Smoothed Particle Hydrodynamics (SPH) kernel:
  \begin{eqnarray*}
    spline(x) = & 1 - (3/2)x^{2}+(3/4)x^{3} & \mbox{ for } 0 < x \le 1 \\  
    spline(x) = & (1/4)(2-x)^{3} & \mbox{ for } 1 < x \le 2 \\
    spline(x) = & 0 & \mbox{ for }  2 < x 
  \end{eqnarray*}

\item Walking from particle to particle, identify local maxima 
  throughout the density field. Apply a first density threshold 
  $\rho_t = 80$ (which roughly corresponds to $b =0.2$ used as standard by the FOF
  algorithm) to all particles, and link particles with a density above $\rho_t$
  to their closest local maximum. These groups of
  particles are defined as (sub)structures.
  
\item Identify saddle points in the
  density field between these groups. Use these saddle points to create branches connecting maxima together, in order to build a 
     structure tree i.e. a hierarchy of nodes where each node
  contains  a collection of particles whose associated density is
  enclosed between two values. The lowest value is the density threshold used to create the first node; the highest is the density associated to the lowest saddle point (if any) detected inside it. The lowest (``first'') level nodes are created by linking groups together whose saddle point is above the first threshold $\rho_t$. The node structure tree is then created by sorting groups in ascending order according to the value of the density associated with their saddle points. 
  
\end{enumerate}

\begin{figure}[ht]
  \centering
  \includegraphics[width=8 cm]{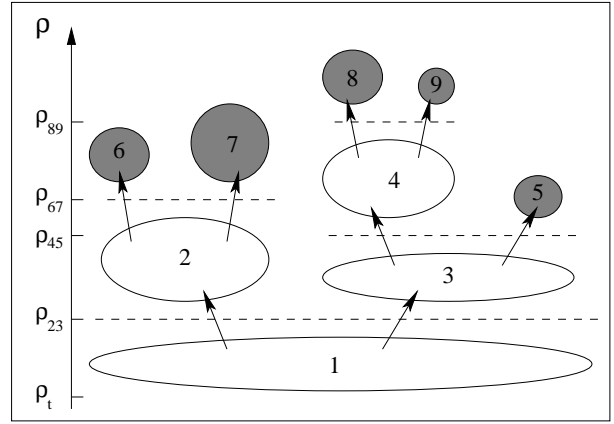}
  \caption{Example of a node structure tree as computed with AdaptaHOP. Ellipses
  are nodes. The arrows show relationship between nodes. In a branch of the
  tree two levels are separated by a saddle point (indicated by a dashed line) in the density
  profile. Leaves in the node structure tree are shown in grey. More massive leaves are represented 
  by larger circles. Density of nodes decreases from top to bottom.}
  \label{ex_nodetree}
\end{figure}

This last item is best explained by
Fig. \ref{ex_nodetree} where the nodes are represented by an ellipse, and
sorted according to their order of creation. The arrows represent how
nodes are linked to one another: the first node to be created in this example is
node 1, with all its particles having a density higher than $\rho_t$; then the
lowest saddle point density is $\rho_{23}$ which separate nodes 2 and 3. 
The particles of node 1, whose density is
greater than $\rho_{23}$ are then split between node 2 and node 3 depending
on how close they are from the density maxima of these 2 nodes. The same
procedure is then repeated to create nodes 4 and 5 from node 3 using a new 
density threshold $\rho_{45}$. This eventually leads to defining nodes 
5, 6, 7, 8, 9 as substructures or ``leaves'' because their particles cannot be
cannot be split anymore between higher level nodes.  

Whereas we can logically define a AdaptaHOP halo as the collapsed node structure tree corresponding
to a group of particles above the $\rho_t$ density threshold, its decomposition into a main halo and a collection of
subhaloes is more tricky. The main problem lies in the fact that nodes are not in general associated with physical objects.
Only end-of-chain nodes, i.e. leaves, have a physical meaning, so we have to re-arrange the node structure tree
in order to build the main halo and the subhaloes.
We tackle this issue in the next subsection (subhalo detection).  

\subsection{Dark matter subhalo detection: One step methods}

These last remarks naturally lead us to address the issue of substructure identification, 
i.e. the detection of subhaloes within haloes as well as subhaloes within subhaloes.  
The main problem we are faced with concerns the node structure built with AdaptaHOP and described 
at the very beginning of the previous section: nodes are not 
in general associated with physical objects. Only the end-of-chain nodes,
i.e. the leaves, have a physical meaning as they are the only true local density maxima in
the density profile of their host halo. Since we need to define physical objects as 
subhaloes, a method is needed to create a tree comprised of a main halo and its
subhaloes from the node structure tree computed by AdaptaHOP. We propose several 
such methods in this section, and compare/contrast their advantages and disadvantages
in the next (Merger Histories).

The obvious choice would be to define as
subhaloes all the leaves in the node structure tree. Still referring to the
example shown in Fig. \ref{ex_nodetree}, this means that we would define nodes 5, 6, 7, 8, 9 as
subhaloes (shown in grey in this figure) and associate nodes 1, 2, 3 and 4
(in white) to the main halo. However, this method is not very satisfying because
it leads to the loss of the hierarchy of subhaloes: we would be left with only 2 levels 
of structures, making it impossible to account for the presence of a subhalo within another subhalo.
Moreover not all density maxima can be defined as subhaloes, as one naturally expects
the main halo itself to be centered on a density maximum.

For this reason, one is forced to lay down two simple, intuitive rules to build a halo tree:
\begin{enumerate} 
\item The main halo and each of the subhaloes of the halo tree must contain one {\em unique} leaf 
from the node structure tree. 
\item The hierarchy between nodes of the node structure tree must be turned into a hierarchy 
of halo, subhaloes, sub-subhaloes etc ... i.e. when a node contains two 
leaves, one of the leaves is defined as the subhalo and the other one, together with the node, 
forms the host (sub)halo of this subhalo. 
\end{enumerate}

Even though these two rules are enough to define a halo tree, they do not
by any means, guarantee the unicity of the solution. In order to remain consistent 
with the basic principle of detection algorithms, we only consider methods 
involving particle positions to build our halo trees in this paper. 
In other words, we only authorise ourselves to distinguish between (sub)structures based on 
two physical quantities: density or mass.

\begin{figure}[ht]
  \centering
  \includegraphics[width=8 cm]{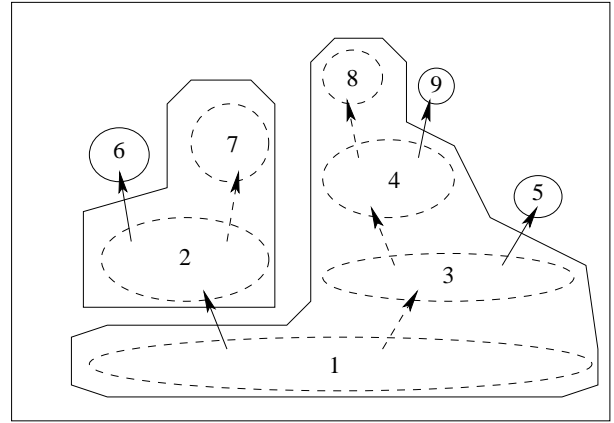}
  \caption{Halo tree obtained from the node structure tree
  shown in Fig. \ref{ex_nodetree} using the DPM method. Haloes and subhaloes are created by 
  collapsing several nodes together. More massive leaves are represented by larger circles.
  Density of nodes decreases from top to bottom. Nodes and connections removed from the node structure
  tree are shown with dashed lines.}
  \label{ex_halotree}
\end{figure}

Arguably, the most natural thing consists in building this halo tree by collapsing  
the node structure tree along a branch containing the most dense leaf:
the particles contained in this leaf are (arbitrarily) chosen to be part 
of the main halo itself, along with all the particles contained in
the lower node levels in which the leaf is included, until we reach the first node (lowest level). 
We then define subhaloes by repeating this procedure for the second most dense leaf, 
and so on and so forth, until all leaves have been accounted for. 
We call this method the Density Profile Method (DPM)  
and illustrate how it works in Fig. \ref{ex_halotree}, which is to be compared
to Fig. \ref{ex_nodetree} depicting the original tree node structure. 
As node 8 is the leaf with the highest density, and is spatially included
in node 4, itself being included in node 3, itself being included in 
node 1, we simplify the tree node by collapsing branch 8-4-3-1 into a single, main halo 
($1\cup3\cup4\cup8$), and replace solid lines and arrows around nodes 8,4,3 with dashes  
in Fig. \ref{ex_halotree} to mark that their particles are now part of a unique object.  
Note that the centre of the main halo is therefore located in
leaf 8. Moving on to the second highest density leaf 9, we find that it is
now only included in node 1 since nodes 3 and 4 have been removed, and therefore 
we have to count it as a subhalo of halo ($1\cup3\cup4\cup8$).
Leaf 7 comes next, which is contained in node 2 which is itself included 
in the main halo: we therefore collapse the branch starting from leaf 7 into
a unique subhalo ($7\cup2$) of halo ($1\cup3\cup4\cup8$). Leaf 6 is spatially included in this new ($7\cup2$)
subhalo, it is therefore a subhalo of this subhalo. Finally, leaf 5 
stands alone in halo ($1\cup3\cup4\cup8$) since we removed node 3: it becomes yet another of its subhaloes.
It is obvious that this solution follows the two rules we layed out for
defining haloes and subhaloes as objects with different levels in the
structure tree because each of the main halo ($1\cup3\cup4\cup8$) on level 1, 
its three subhaloes (9,$7\cup2$,5) on level 2, and the unique sub-subhalo (6) 
on level 3, contain a single leaf from the original node tree. 

Moreover, from this simple example, we can easily understand how changing the criterion to pick the 
first leaf --- for example picking the most massive one instead of the most dense ---, we would have defined 
$1\cup2\cup7$ as the main halo , and found it had two subhaloes, $3\cup4\cup8$, and 6. Finally, instead of 
having node 6 as the only level 3 structure of the halo structure three, both leaves 5 and 9 would be 
subhaloes of subhalo $3\cup4\cup8$. We call this second method the Most massive Sub-node Method (MSM).
A fundamental difference between halo trees built using DPM and MSM is that the centre of 
the main halo is now located in leaf 7 (MSM) instead of leaf 8 (DPM).
We also emphasise that the criterion used to group nodes to form 
haloes and subhaloes has an effect not only on the mass of the main halo but on the 
hierarchy (level number) of subhaloes as well. 

\subsubsection{Individual examples}

Both methods are then run on the same $N$-body simulation 
described in paper GalICS I \citep{Hatton2003}. This simulation 
contains $256^{3}$ particles of 8.03 $10^9$ M$_{\odot}$ enclosed in a volume
of 150 comoving Mpc on a side, with periodic boundary conditions. Values for 
cosmological parameters are given in Table\ref{table:simu}.
  
\begin{table}[hbp]
  \caption{Simulation details}    
  \label{table:simu}
  \centering
  \begin{tabular}{c c}
    \hline
    Number of bodies & $256^3$ \\ 
    Particle mass & 8.03 $10^9$ M$_{\odot}$\\
    Box size & 150 Mpc\\
    Omega matter $\Omega_M$ & 0.333\\
    Omega lambda $\Omega_{\Lambda}$ & 0.667\\
    Hubble parameter h & 0.667\\
    $\sigma_8$ & 0.880\\
    initial redshift & 35.6\\
    \hline
  \end{tabular}
\end{table}

We now present two example of haloes using the MSM method to define halos and their
hierarchy of subhaloes. Our goal is to check by eye  (arguably the best tool
to do the job) that we can detect all the subhaloes.

\begin{figure}[ht]
  \centering
  \includegraphics[width=9 cm]{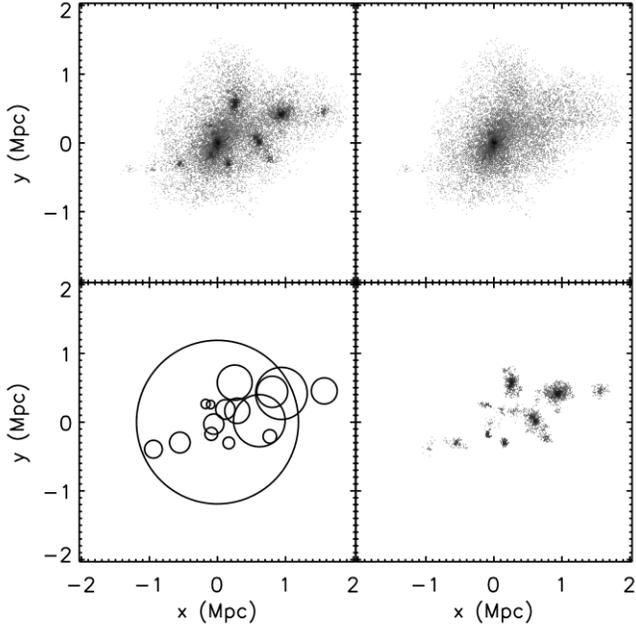}
  \caption{Top left: original AdaptaHOP halo with subhaloes, top right: main (MSM) halo (without
    subhaloes), bottom right: subhaloes, bottom left: circles marking the ``virial'' region of the
    halo and its subhaloes.}
  \label{halo_MSM_48}
\end{figure}

\begin{figure}[ht]
  \centering
  \includegraphics[width= 9 cm]{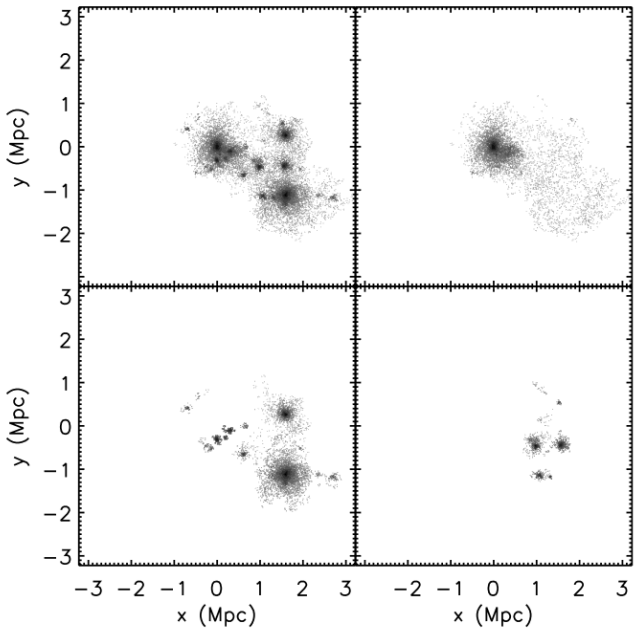}
  \caption{Top left: original AdaptaHOP halo with subhaloes, top right: main (MSM) halo (without
    subhaloes), bottom left: subhaloes of the main halo, bottom right: subhaloes of the subhaloes.}
  \label{halo_MSM_1421}
\end{figure}

As shown in Fig. \ref{halo_MSM_48}, the MSM can detect subhaloes quite
well, for a quite relaxed halo of mass 1.2 $10^{14}$ M$_{\odot}$ with its 15
subhaloes, and of mass $10^{14}$ M$_{\odot}$ without them.
It is able to neatly remove all 
subhaloes as shown in the top right panel: no spurious holes appear in the 
density field where these subhaloes have been removed. The main halo (top right panel) 
has a smooth profile even though the
subhaloes shown on the bottom left panel come in various shapes and sizes. 
Most of the subhaloes shown here have the main halo as
host. It is easier to observe the size of the various subhaloes by drawing their
``virial'' region (defined here as a sphere centered on the centre of the halo or
subhalo and whose radius is such that the average density is $<\rho_r> =
\rho_{200} = 200 \times \rho_c$) which are shown on the bottom left panel.

More revealing is how MSM performs when the halo is still perturbed by a major
merger with another object (i.e. when the mass ratio between the
two haloes is higher than 1:3). Such a case is 
detailed in Fig. \ref{halo_MSM_1421}. 
The mass of the AdaptaHOP halo (with subhaloes) is 1.2 $10^{14}$ M$_{\odot}$, but this time
the mass of the main (MSM) halo is only 5.7  $10^{13}$ M$_{\odot}$, i.e. its 23 subhaloes
contain slightly more mass than it does. Moreover, since one of the subhaloes is about the 
same size as the main halo we can expect it to contain quite a few subhaloes itself.
Indeed, we are not disappointed: as shown in the top right panel of the figure, the main halo 
even harbours a plume of low density particles 
whose origins probably lie in the tidal stripping of material between the bigger
subhalo and the host halo. As expected the biggest subhalo hosts a few subhaloes of its own,
which are large and dense enough to be seen in the bottom right panel of Fig.
\ref{halo_MSM_1421}. Usually level 3 subhaloes are much smaller and more difficult to 
spot on these kind of plots, but the main conclusion is that the MSM method seems 
to perform well independently of the state of relaxation of the AdaptaHOP halo. 

\subsubsection{Density vs mass criteria}

At this point, we have to decide which of the two methods is the best 
suited to reach the goal that we set for ourselves at the beginning of this paper: to build 
the most reliable (sub)halo merging history tree possible. Is it the DPM, where each subhalo has a lower 
peak density than its host? Or is it the MSM where each subhalo is less massive than 
its host?
We now compare these two methods. 

\begin{figure}[bp]
  \centering
  \includegraphics[width=8 cm]{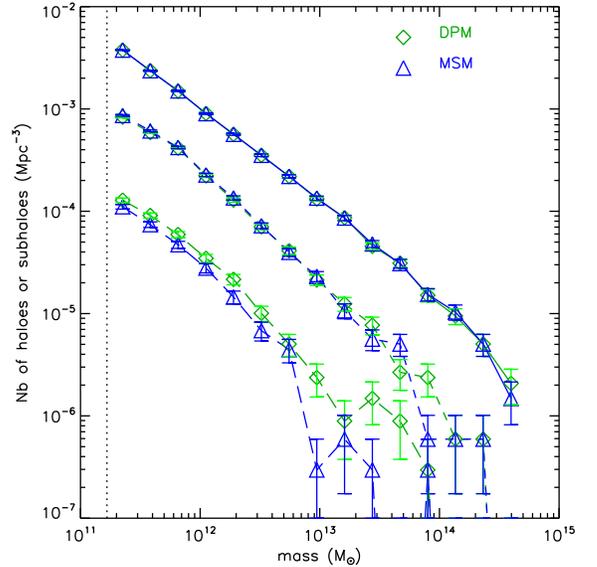}
  \caption{For both methods DPM (diamonds) and MSM (triangles), the mass functions of haloes, 
subhaloes of haloes and subhaloes of subhaloes are plotted with solid, short-dashed and long-dashed curves
respectively. 
The error bars correspond to Poisson uncertainty.
The vertical dotted line corresponds to the 20 particles detection threshold.
}
  \label{DPM_MSM_massfunction}
\end{figure}

The mass functions obtained for both DPM (diamonds) and MSM (triangles) methods at $z=0$ are 
represented in Fig. \ref{DPM_MSM_massfunction}. For each method, the mass function of main haloes 
(level 1 structures), subhaloes of haloes (level 2 structures) and subhaloes of subhaloes (level 3 and 
above structures) are represented respectively with solid, short-dashed, and long dashed curves respectively. 
From the figure it is apparent that the mass functions of haloes are very similar in both methods, with small 
differences barely perceptible to the naked eye. These differences are only due to the fact that the mass of haloes can vary from one method to the other, when the most massive and the most dense leaves do not coincide. So even 
though the total number of haloes detected at redshift 0 is the same, the number of main haloes per mass bin is expected
to vary slightly. The same comment holds for the mass function of subhaloes, except at the high mass end where 
the difference becomes noticeable and it seems that slightly less subhaloes of halos are detected with the 
MSM method. Differences between methods only become apparent in the mass 
function of subhaloes of subhaloes (so called level 3 and above). Among the 9124 subhaloes detected with both methods
we count 2170 subhaloes of subhaloes: 1201 subhaloes detected with the DPM method and 969 with the MSM method. 
We believe that these numbers are large enough to trust that the trend of the DPM always producing a higher number of subhaloes 
of subhaloes is real. This is especially true since it is verified
for every mass bin, even if the number of these subhaloes is small for masses above 7.7 $10^{12}$ 
M$_{\odot}$, (18 for the DPM, 5 for the MSM), and from these bins alone we would be hard pressed 
to draw any conclusion.

\begin{figure}[tp]
  \centering
  \includegraphics[width=8 cm]{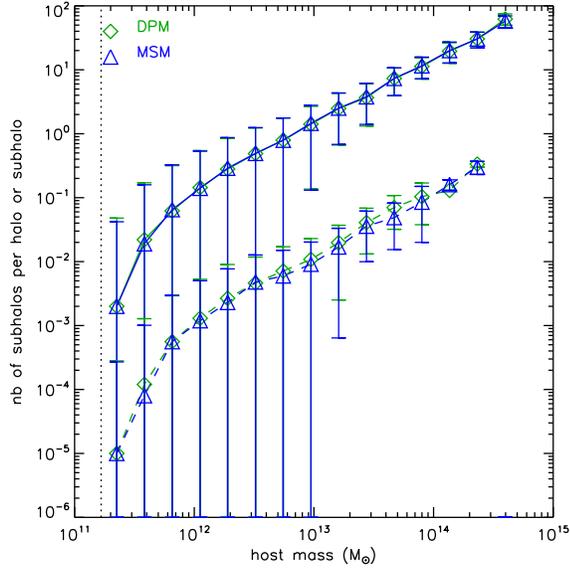}
  \caption{For both methods DPM (diamonds) and MSM (triangles), the average number of subhaloes 
per halo (plain curve), and subhaloes per subhalo (dashed curve vertically translated by 2 dex downwards 
for clarity) are shown.
The error bars correspond to the mean quadratic dispersion.
The dotted vertical line corresponds to the 20 particle detection threshold.}
  \label{nsub_per_halo_DPM_MSM}
\end{figure}

The number of subhaloes per halo or subhalo is another aspect of substructure selection
methods that is worth quantifying. This data is shown in Fig.
\ref{nsub_per_halo_DPM_MSM}. For each method (DPM marked with diamonds and MSM with triangles), we show here the average number 
of subhaloes per halo (plain curves) and the number of subhaloes per subhaloes (dashed curves vertically
translated 2 dex downwards for clarity). 
Once again, the average number of subhaloes per halo is nearly the same when using either the DPM or MSM method. As
expected, this number is very low $10^{-3}$ at 2 $10^{11}$ M$_{\odot}$. Haloes of these masses are close to the 
detection threshold and are unlikely to host any subhaloes. In both cases, we reach an average number of at least 1 subhalo per halo from 
$10^{13}$ M$_{\odot}$ onwards. Close to $10^{14}$ M$_{\odot}$, we obtain on average 12 subhaloes per halo. 
The number of subhaloes per subhalo follows the same trend, but the difference between the two methods 
is more marked in that case, with DPM subhaloes hosting slightly more subhaloes than MSM subhaloes. 
This result confirms the impression we got from the mass function (Fig. \ref{DPM_MSM_massfunction})
that the number of subhaloes of subhaloes is indeed higher with the DPM method.   

\begin{figure}[ht] 
  \centering
  \includegraphics[width=8 cm]{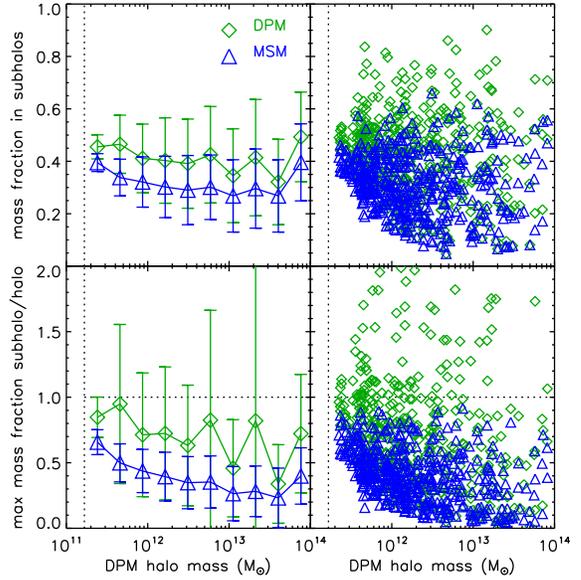}
  \caption{For each halo under $10^{14}$ M$_{\odot}$ that differ in the DPM (diamonds) and MSM (triangles) method, two mass fractions were computed. The first mass fraction is the total mass found in subhaloes over the total mass of the halo (the main halo plus its subhaloes). This fraction is shown in the top panels. The second mass fraction is the mass of the largest subhalo over the mass of its host main halo. This fraction is shown in the bottom panels. The average results are shown on the left panels, the results for each halo are shown in the right panels.
The error bars correspond to the mean quadratic dispersion.
The dashed vertical line corresponds to the 20 particle detection threshold.}
  \label{nfsub_DPM_vs_MSM}
\end{figure}

Coming back to the AdaptaHOP haloes (the haloes originally detected by AdaptaHOP before subhaloes are removed
from them), which, unlike the main haloes are the same for both methods, we select those for which the main 
halo has a different mass in each method and compute the mass fraction 
contained in all its subhaloes. By definition this mass fraction is always smaller than 1 since an AdaptaHOP halo
contains all its subhaloes. The result of this exercise is presented in the top panels of Fig.
\ref{nfsub_DPM_vs_MSM}, where the average mass fraction for the DPM (diamonds) and the MSM methods (triangles) 
(left panel) and the individual values (right panel) are shown. 
The mass fraction found in DPM subhaloes is close to 0.5 for the smallest haloes and decreases down to 0.3 near 
$10^{13}$ M$_{\odot}$, to rise again and reach 0.5 at $10^{14}$ M$_{\odot}$. The mass fraction in MSM subhaloes follows
 the same trend but the mass fraction is always about 0.1 lower. The error bars are smaller as well for the MSM results
and this can be viewed more clearly on the top left panel: the scatter in mass fraction is more pronounced for the 
DPM method.
When using the DPM or MSM method on an AdaptaHOP halo, we obtain the same number of subhaloes, which is the number 
of density maxima (leaves in the node structure tree) minus one that is defined as the center of the main halo itself.
For each of these AdaptaHOP haloes the number of subhaloes therefore is initially the same. However, when the DPM 
and MSM methods differ on the choice of the center leaf, a higher mass fraction of the AdaptaHOP halo is
 found in DPM subhaloes than in MSM subhaloes, which is somewhat expected since MSM picks the most massive 
center leaf to define it as part of the main halo. What is more worrisome, is that looking at 
the bottom panels of Fig. \ref{nfsub_DPM_vs_MSM}, we realise that the mass ratio between a subhalo
and the main halo can be greater than one with the DPM method across the entire mass range spanned by the 
$N$-body simulation. Among the 33718 DPM main haloes with a mass lower than $10^{14}$ M$_{\odot}$, 455 differ 
when using the MSM method. Among these, 93 have a subhalo-halo mass ratio greater than 1 and 35 greater than 1.5.
This never happens for MSM haloes, 
which have a maximum subhalo-halo mass ratio lower than 1 by construction. In the left panel, we see that this mass 
fraction is on average lower than 0.7 for the MSM haloes and much smoother than the DPM curve, with a reduced 
scatter. It decreases from 0.7 for the smallest haloes to around 0.2 at 2 $10^{13}$ M$_{\odot}$, and rises again 
to reach 0.5 for haloes with masses close to $10^{14}$ M$_{\odot}$. 

The DPM method (density criteria) and the MSM method (mass criteria) do not differ 
in most cases as the most massive leaf in the node structure tree generally is the densest as well. However,
 when they do differ, the hierarchy of subhaloes is modified (we obtain more subhaloes in subhaloes 
with the DPM method) and the mass ratio between a subhalo and its host halo is also modified.
This latter can reach unphysical (greater than unity) values with the DPM method
which leads us to use the MSM method as our preferred one step method in the rest of this 
paper. 

Having selected a robust (in the sense that it uses an integral quantity, i.e. the mass, rather 
than its spatial derivative, i.e. the density) and intuitively satisfying method (MSM haloes are more massive 
than any of their subhaloes) as well as assessed its ability to describe 
the instantaneous structure of dark matter haloes and subhaloes (i.e. the analysis of an isolated
snapshot of an $N$-body simulation), we now proceed 
to study how it fares at capturing their much more complex time evolution.    

\section{Merger histories}

\subsection{Constructing a merger tree}

Although over the course of a numerical $N$-body simulation most (sub)haloes lose mass and disappear, it is 
customary to refer to the ``growth'' of haloes since these mass losses also feed the formation
of larger objects. Tracking how and when a (sub)halo acquires its mass is called building a
merger history tree for this specific (sub)halo. The further complication with respect to the detection 
of subhaloes previously discussed is that this mass assembly is a dynamical and continuous 
process, which is only captured by a finite number of discrete time outputs (typically 50 between redshifts
20 and 0, see e.g. \citet{Hatton2003}) of an $N$-body simulation. This means that time resolution issues 
superimpose on mass resolution issues that were our sole limitation up to now.

\begin{figure}[bp]
  \centering
  \includegraphics[width=8 cm]{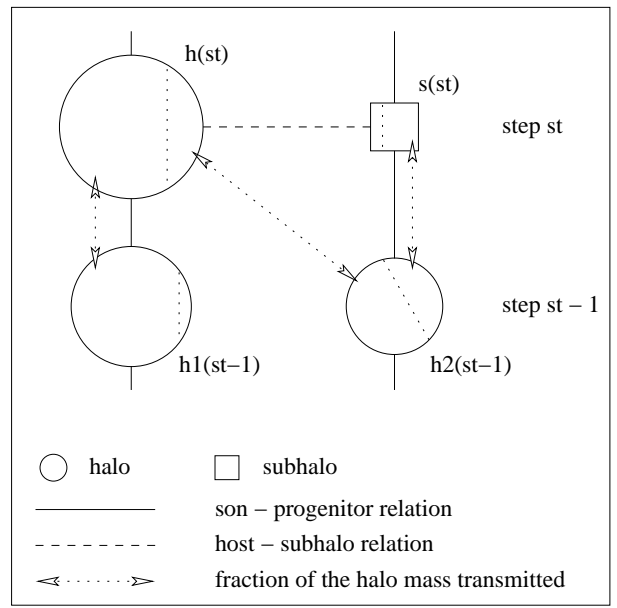}
  \caption{In a simplified tree built as described in the text (Sect. 3.1), a subhalo may appear without seemingly having a  
   progenitor. Whenever possible, the simplified merger tree 
   is therefore modified so that the subhalo $s(st)$ in the figure becomes the main son of halo $h2(st-1)$, 
  even though $h2(st-1)$ has given most of its particles to $h(st)$ and not $s(st)$.}
  \label{sub_prog}
\end{figure}

Once again, all the information available to build the merging trees is carried by particles, and more 
specifically by their belonging to a particular (sub)halo at a given time output $st$ and another one
at the previous/subsequent snapshot. In other words, by monitoring the exchange of particles between
haloes or subhaloes detected in two consecutive snapshots of an $N$-body simulation, one can 
build ``simplified'' merger trees by laying down the following set of rules:
\begin{itemize}
\item Each (sub)halo $i$ at step $st$ can only have 1
  son at step $st+1$, i.e. fragmentation is not taken into account, hence the ``simplified''
  adjective used to describe the merging tree. 
\item Assuming $m_{ij}$ is the common mass between a (sub)halo $i$ of mass $m_i$ at step $st$ and 
  a (sub)halo $j$ of mass $m_j$ at step $st+1$, the son of $i$ is chosen as the (sub)halo $j$ for which 
  $m_{ij}/{m_i}$ is maximal.     
\item Conversely, the (sub)halo $i$ of mass $m_i$ at step $st$ is a progenitor of the (sub)halo $j$
  of mass $m_j$ at step $st+1$ if, and only if $j$ is the son of $i$. The {\em main} progenitor of $j$ 
  is the (sub)halo $i$ for which the ratio $m_{ij}/{m_j}$ is maximal.
\end{itemize}

Merger trees built from $N$-body simulations in a similar way for haloes only 
(with DENMAX: \citet{Roukema1997} or FOF: \citet{Kauffmann1999,Hatton2003,Helly2003,Nagashima2005}),
and more recently for subhaloes as well (using SKID \citep{Okamoto2000} or SUBFIND \citep{Springel2005,De_LuciaBlaizot2007,Fakhouri2008})
are quite common. 
Note that these merger trees are not binary. Each (sub)halo only has one son, but mergers can occur 
between more than two (sub)haloes, depending on time resolution. Also when two haloes merge, if the first one survives as a subhalo
the merger does not appear immediately as a ``real'' merger event (two branches become one) since the son of this first halo is a subhalo which is distinct from 
the son of the second halo which we call the main halo. A merger only really occurs when the subhalo has completely dissolved in the 
main halo at a later time output. 
Finally, a twist to the previous rules, and specific to mergers trees including subhaloes has to be introduced.
The reason for this, as illustrated in Fig. \ref{sub_prog}, is that
often a subhalo, $s(st)$, is not the main son of any halo. Assume for instance that its host halo $h(st)$ has two (or more) progenitors 
$h1(st-1)$, the main one, and $h2(st-1)$. If $h2(st-1)$ has given most of its mass to $h(st)$, but at the same time $s(st)$ got most of its particles from $h2(st-1)$
then $s(st)$ is an orphan according to the rules. We therefore choose, in these cases, to modify the merger tree so that $s(st)$ becomes the main son of $h2(st-1)$ 
and $h2(st-1)$ the main progenitor of $s(st)$. In short, this modification accounts for the fact that when 
two haloes merge, the smallest one can lose most of its mass to the newly identified main halo while still retaining a clear identity by becoming a subhalo.  
Another, arguably better, way to proceed would be to use the most bound particle(s) to define which of the possible sons should be the main one (e.g. \citet{Okamoto2000}). 

In the remaining subsections of the paper we will be talking about tree branches. What we define as a branch is a succession of haloes and subhaloes 
linked together between two outputs of an $N$-body simulation by a main progenitor -- main son relationship. It means that there is only one (sub)halo 
per branch at any given step, and that a branch starts with a (sub)halo which has no progenitor and ends (i) when a merger occurs with another branch or 
(ii) when a (sub)halo has no son at the next output or (iii) when the current output is the last one (e.g. $z=0$). 

\subsection{Example of a merger tree built with and without subhaloes.}

\begin{figure}[ht]
  \centering
  \includegraphics[width=8 cm]{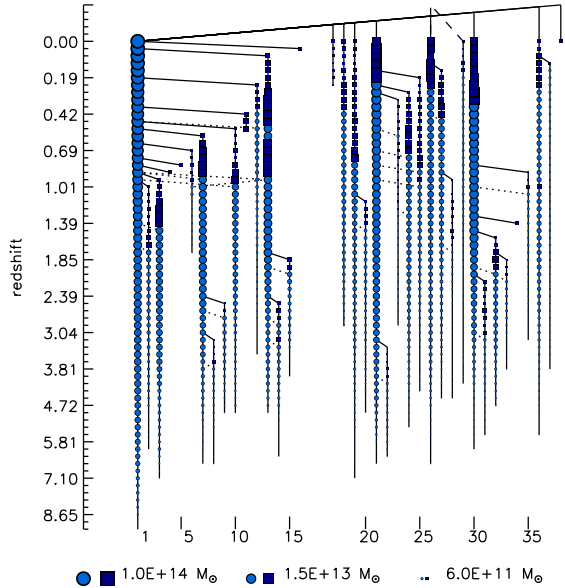}
  \caption{Example of a merger tree computed with MSM subhaloes according to the rules described in the text (Sect. 3.1).
    Circles represent main haloes, squares subhaloes. The main halo at
    redshift 0 is the one shown in Fig. \ref{halo_MSM_48}. Its merger tree is shown in the left hand side 
    of the figure, whereas the merger trees of its subhaloes are shown on the right hand side. Main haloes are connected to their subhaloes 
   using horizontal solid lines and mergers between (sub)haloes are indicated by horizontal dotted lines.
   A mass threshold was applied not to show the less massive (thick) branches and limit this tree to less than 50 branches. Only the 40 most massive branches of the full merger tree are shown in this figure.}
  \label{merger_tree_48_MSM}
\end{figure}

\begin{figure}[ht]
  \centering
  \includegraphics[width=8 cm]{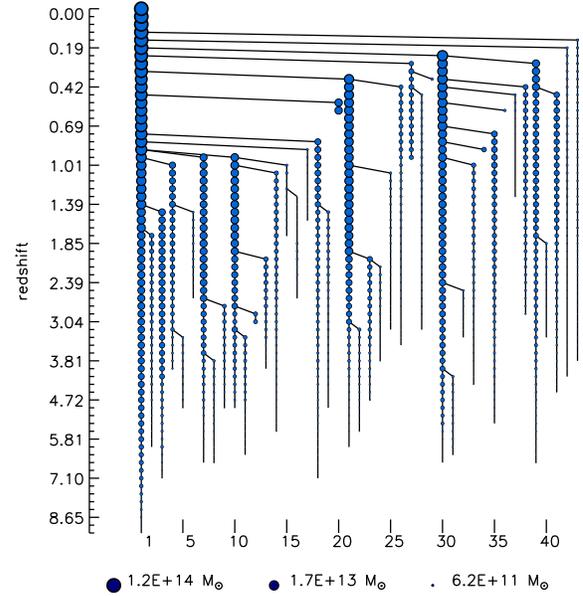}
  \caption{Same as Fig. \ref{merger_tree_48_MSM} except that the subhaloes have not 
    been taken into account (i.e. it is the merger tree of an AdaptaHOP halo which includes
    but does not separate subhaloes). A mass threshold was applied so as not to show the less massive (thick) branches and limit this tree to less than 50 branches. Only the 42 most massive branches of the full merger tree are shown in this figure.}
  \label{merger_tree_48_HOP}
\end{figure}

Figure \ref{merger_tree_48_MSM} shows the full merger tree of the halo shown in Fig. \ref{halo_MSM_48}, with subhaloes determined 
via the MSM, albeit for this particular halo, little difference would occur if this merger tree was constructed with the DPM. 
In this figure, haloes are represented as circles, subhaloes as squares. The main halo and its subhaloes are on the first line, 
and all of their progenitors on subsequent lines as time flows from the bottom to the top of the figure. 
Each column is a branch of the tree either linked to the halo or one of its subhaloes. The main progenitor is always in the same 
column as its son. When a merger occurs one branch ends (no more objects in this column) 
and a line connects it to the branch it has merged with. In this plot we show that 16 branches directly lead to the main halo at redshift 
0. The first one is the main branch (the trunk), and the 15 other ones which end before redshift 0 are called secondary branches. 
The other 22 ``sub''-branches define the merger histories of subhaloes hosted by the main halo at redshift 0. The relationships between haloes 
and subhaloes are indicated by a line at the top of each branch. 
They show whether a subhalo is hosted by the main halo itself (solid lines connecting the two objects) or, as for the branch in Col. 29, 
another subhalo (dashed lines connecting the two objects). Some lines linking branches are dotted: these mark mergers between (sub)haloes which 
resulted in both objects retaining their identities (one becomes the subhalo of the other).
When this happens either of two things can occur at later times: (i) the branch of the subhalo merges with the branch of its host or (ii) the
subhalo becomes a stand alone halo again. Both cases are present in the halo merger tree of Fig. \ref{merger_tree_48_MSM}, with case (i) being more 
frequent (branches in Cols. 2, 3, 7, 8, 9, 10, 15, 27, 28, 31, 32, 33, 35) than case (ii) (branches in Cols. 13, 14, 23, 25, 34). Physically, case (i) 
corresponds to progressive mass stripping of the subhalo through dynamical friction, and case (ii) 
to structures which fly by one another several times on elongated orbits before merging together for good.

If we do not keep track of the subhaloes, as shown in Fig. \ref{merger_tree_48_HOP} where we plot the merger tree of the AdaptaHOP 
halo (the halo which includes the main halo and all subhaloes) the same phenomenon of multiple fly-bys before merger leads to branches being 
cut into pieces. The lower part of the branch will merge with the halo at the first encounter, and the top part of the branch will reappear as a new 
branch each time the ``subhalo'' jumps down a level in the halo tree, e.g. becomes a halo again if it had become a ``level 2 subhalo'' during the first encounter (branches in Cols. 18 and 20).   
However, the general impression one gets from glancing at Fig. \ref{merger_tree_48_HOP} is that the AdaptaHOP halo merger tree is quite close to the MSM one  
(Fig. \ref{merger_tree_48_MSM}). More specifically, if we ignore the squares and replace the dashed lines by solid lines in this 
figure it is clear that many mergers observed in Fig. \ref{merger_tree_48_HOP}, are also found in Fig. \ref{merger_tree_48_MSM} (branches in Cols. 2 and 3 are identical,
branch 4 in Fig. \ref{merger_tree_48_HOP} is branch 13 in Fig. \ref{merger_tree_48_MSM}, and so on and so forth...).

One of the main ideas behind this paper is to construct ``well behaved'' merging trees from $N$-body simulations, on which we can graft semi-analytic 
models (SAM: e.g. \citet{Roukema1997,Kauffmann1999,Hatton2003}) of galaxy formation and evolution on top of them. In order for this to be possible, we 
expect our trees to be devoid of certain features. For instance, subhaloes should only appear after a merger involving main haloes has taken place. 
This means that any subhalo should have at least one progenitor. We then expect a subhalo to be either stripped of its mass and merge with its
host structure, or become temporarily distinct again (see discussion above). Which means that, in principle, it should not be possible for a subhalo 
to ever be part of a main branch. A thorough analysis of our merging tree building method(s) therefore leads us to define three types of anomalies that 
are to be avoided, or at least reduced to a minimum of occurrences:

\begin{enumerate}
\item Anomaly of the first kind: a subhalo has no progenitor. 
\item Anomaly of the second kind: a subhalo merges with its host but its branch does not end.
\item Anomaly of the third kind:  a subhalo swaps identity with its host.
\end{enumerate}

\begin{figure}[ht]
  \centering
  \includegraphics[width=8 cm]{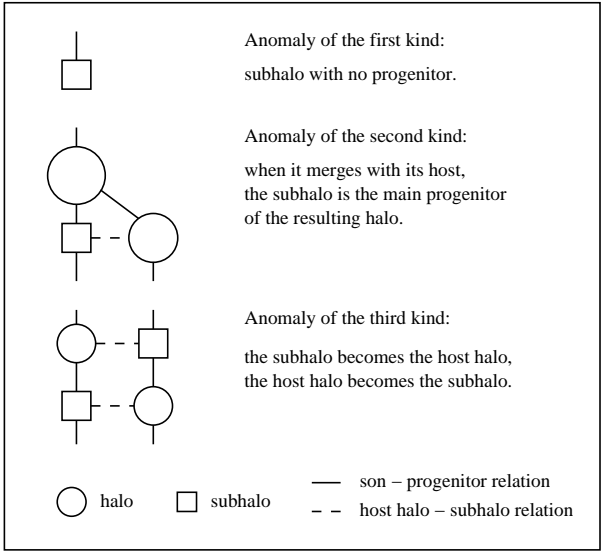}
  \caption{The three main anomalies that one can run into when building a merger tree with subhaloes (see text, Sect. 3.2, for details).}
  \label{err_tree}
\end{figure}

These are depicted in Fig. \ref{err_tree}. Measuring the occurrence of these anomalies within 
merger trees built with different methods is a good way not only to assess the relative performance
of these methods at capturing the complex dynamics of the halo merging process, but also to estimate the 
suitability of these merger trees to be used as backbones for SAMs. We now proceed to perform this comparison.  

\subsection{A two time step method: the Branch History Method (BHM)}

\begin{figure}[ht]
  \centering
  \includegraphics[width=8 cm]{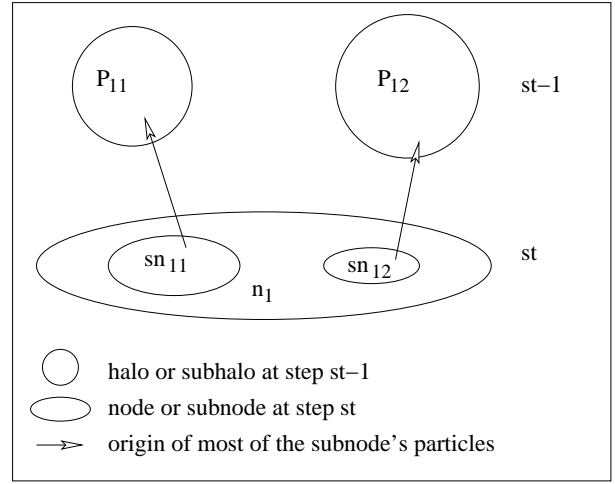}
  \caption{Illustration of the BHM method (see text, Sect. 3.3, for details), where a node, $n_{1}$, and two of its subnodes, 
    $sn_{11}$ and $sn_{12}$ at step $st$, 
    are linked to the progenitors of subnodes (either haloes or subhaloes) at step $st-1$ (named $P_{11}$ and $P_{12}$).}
  \label{node_prog}
\end{figure}

The occurrence of anomalies in the merger history of haloes is caused by the (poor) finite time resolution necessarily used 
to store the wealth of information contained in $N$-body simulations. It therefore seems natural to try to include information
coming from various time outputs to get rid of them. In this section, we present a method that uses information over two 
such time outputs, but it can in principle be extended beyond this (at the expense of computational power and complexity)
to build truly ``perfect'', i.e. virtually anomaly free, trees.
Like the other methods, the Branch History Method (BHM) starts with the node structure tree computed with AdaptaHOP, but
it works in the following way: 
\begin{itemize}
\item Load the (sub)halo distribution of the previous time output in memory. 
      If there is none (first snapshot), use the MSM method to detect (sub)haloes.
\item Construct the halo tree from the highest level of the node structure tree to the lowest one (which is the main 
 halo itself: see Sect. 2 ``structure detection'' for details) for the current time output. 
\item When a node $n_1$ contains two subnodes $sn_{11}$ and $sn_{12}$, check that the mass of $(n_1 + sn_{11})$ is greater that the mass of $sn_{12}$.
 Similarly when using $sn_{12}$ instead of $sn_{11}$, check that the mass of $(n_1 + sn_{12})$ is greater that the mass of $sn_{11}$, to ensure 
 that if one of the subnodes is defined as a subhalo,  its mass will be lower than that of its host.
\item Compute the main progenitor of each of the objects $sn_{11}$, $sn_{12}$, $(n_1 + sn_{11})$ and $(n_1 + sn_{12})$, i.e. the halo or subhalo with which these objects 
 have most mass in common at the previous time output. We shall name those progenitors $P_{11}$,$P_{12}$,$P_{1+11}$ and $P_{1+12}$. See Fig. \ref{node_prog} for illustration.
\item Apply the following criteria:
  \begin{enumerate}
    \item If $P_{11} = P_{1+11}$ and $P_{12} \neq P_{1+12}$, then $sn_{12}$ is a subhalo and $(n_1 + sn_{11})$ its host. Vice versa if $P_{11} \neq P_{1+11}$ and $P_{12} = P_{1+12}$, $sn_{11}$ is a subhalo and $(n_1 + sn_{12})$ its host.
    \item If $P_{11} = P_{1+11}$, $P_{12} = P_{1+12}$, and $P_{11}$ and $P_{12}$ are both haloes, use the same criteria as the 
     MSM method: if the mass of $sn_{12}$ is lower than the mass of $sn_{11}$, then $sn_{12}$ is a subhalo and $(n_1 + sn_{11})$ its host.
    \item If $P_{11} = P_{1+11}$, $P_{12} = P_{1+12}$, $P_{11}$ is a subhalo and $P_{12}$ a halo then $sn_{11}$ is a subhalo and $(n_1 + sn_{12})$ its host. Vice versa if $P_{11}$ is a halo and $P_{12}$ a subhalo then $sn_{12}$ is a subhalo and $(n_1 + sn_{11})$ its host: i.e. make sure that a subhalo remains a subhalo whenever possible.
    \item If $P_{11} \neq P_{1+11}$ and $P_{12} \neq P_{1+12}$, then both $sn_{11}$ and $sn_{12}$ are subhaloes and $n_1$ is their host: the masses of
     $sn_{11}$ and $sn_{12}$ are small compared to that of $n_1$, so there is no reason to decide that one subnode is a subhalo and not the other.
  \end{enumerate}
\end{itemize}

Note that the method we have just outlined differs from that used by \citet{Springel2005} as these authors only take 
advantage of information 
from previous/subsequent time steps to build their merging trees, {\em not} to decide on particle (sub)halo
appartenance. We now proceed to check all merging trees built using the BHM and both one step methods DPM and MSM, 
for the anomalies defined in the previous subsection.

\begin{figure}[ht]
  \centering 
  \includegraphics[width=9 cm]{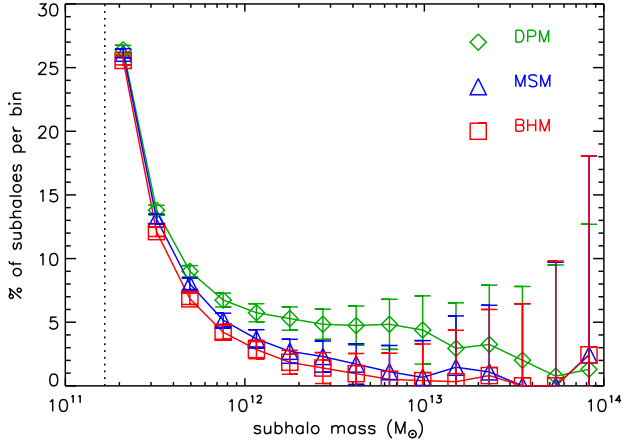}
  \caption{Percentage per mass bin of subhaloes without progenitor, i.e. of the occurrence of the first anomaly, for the three merging tree building methods described 
in the text: DPM (diamonds), MSM (triangles), BHM (squares).
   The error bars correspond to Poisson uncertainties.
   The vertical dotted line corresponds to the 20 particles detection threshold.}
  \label{suborigins}
\end{figure}

Figure \ref{suborigins} shows results obtained for the anomaly of the first kind, i.e. the number of subhaloes for which no progenitor could be assigned. 
These results follow the same trend as a function of subhalo mass whatever the tree building method (DPM, MSM, BHM) used. 
Up to 26 \% of the smaller subhaloes do not have a progenitor but this fraction quickly decreases as the mass of the subhalo increases,  
falling below the 10 \% mark for DPM, MSM and BHM subhaloes more massive than 5 $10^{11}$ M$_{\odot}$, 4.5 $10^{11}$ M$_{\odot}$ and 4 $10^{11}$ M$_{\odot}$ respectively.
For the DPM, this percentage stays between 5.3 and 4.3 \% from 1.4 $10^{12}$ M$_{\odot}$ to 1.2 $10^{13}$ M$_{\odot}$, then decreases below the 3 \% mark from 2.8 $10^{13}$ M$_{\odot}$
onward.
 For the MSM and BHM methods the 5 \% level is reached at 9.2 $10^{11}$ M$_{\odot}$ and  5.6 $10^{11}$ M$_{\odot}$  and the 2 \% mark at 3.3 $10^{12}$ M$_{\odot}$ and 1.4 $10^{12}$ M$_{\odot}$.
In the last mass bin, all methods detect only 1 subhalo with no progenitor, but since the number of subhaloes is higher in this bin for the DPM (77 compared to 41 for MSM and BHM)
the percentage is slightly lower for this method. 

We conclude that as far as anomalies of the first kind are concerned, both the MSM and BHM yield significantly better results, as
their fraction of subhaloes without progenitors is systematically a few \% lower than with the DPM. 
The fact that 26 \% of the smallest subhaloes have no progenitor is not too worrying, as these 
subhaloes are more prone to contamination by Poisson noise: they mostly 
disappear from a time output to the next simply because they lose a few particles and drop below the detection threshold 
of 20 particles. For the larger subhaloes, the main reason for the anomaly to occur is that a subhalo with a similar maximal
 density than its host and coming close to the center of the latter on a radial orbit can be blended with it \footnote{A great advantage of 6D detection is for instance that 2 structures colliding with each other, that would be indistinguishable in 3D at the moment of crossing remain well separated in phase-space.}. At the time output
when the blending occurs, one then detects just one structure: the host. This
anomaly can be understood as a fly-by process occurring for subhaloes, also
occurring with the SUBFIND algorithm and detailed by \citet{Wetzel2009}.

However, at the next time output this need not
be the case, and since we do not allow a (sub)halo to have more than one son, a structure is left without a progenitor. 
Depending on the method one uses to define (sub)haloes, this can happen more (DPM) or less (BHM) often, as different haloes
can be picked as main progenitors.

\begin{figure}[bp]
  \centering
  \includegraphics[width=9 cm]{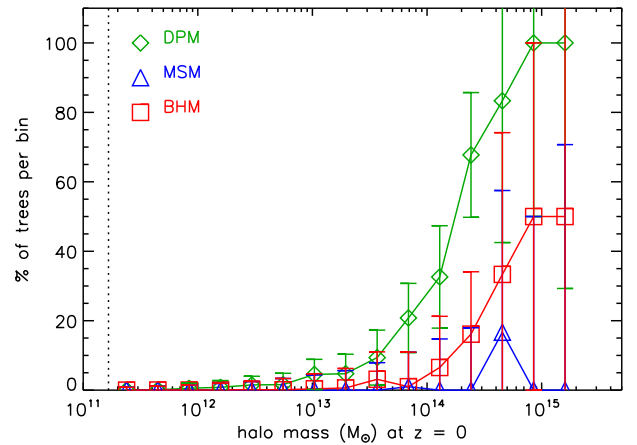}
  \caption{For the three merging tree building methods DPM (diamonds), MSM (triangles), BHM (squares), haloes at redshift 0 
have been sorted into 15 mass bins. 
 The merger trees of each of these haloes were analysed, to detect occurrences of the anomaly of the second kind, 
i.e. a subhalo merging with its host halo but becoming the main 
 progenitor of the resulting main halo. For each mass bin, the percentage of trees in which this anomaly occurred at least once 
is given. 
The error bars correspond to Poisson uncertainties.
The vertical dotted line corresponds to the 20 particles detection threshold.}
  \label{trees_err2}
\end{figure}

\begin{figure}[tp]
  \centering
  \includegraphics[width=9 cm]{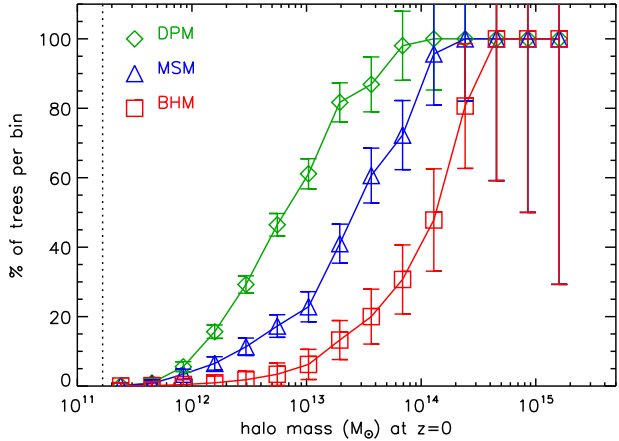}
  \caption{Same as Fig. \ref{trees_err2} but for anomalies of the third kind, i.e. from one step to another, a subhalo is detected in its host halo branch, and the host halo is detected in the subhalo branch. In each bin the number of trees where this anomaly occurred at least once is displayed. The vertical dotted line corresponds to the 20 particles detection threshold.}
  \label{trees_err3}
\end{figure}

In Figs. \ref{trees_err2} and \ref{trees_err3}, we searched the merger trees of each $z=0$ halo built using all three methods
for occurrences of anomalies of the second and third kind respectively. These trees were sorted into 15 mass bins corresponding 
to the mass of the final halo and its subhaloes at redshift 0, meaning that for all method any tree is in the same bin. 

Contrary to the first kind of anomaly, the second kind is quite scarce. It occurs only in 219, 4 and 31 trees for the DPM, MSM and BHM respectively.
Looking at the DPM results in Fig. \ref{trees_err2}, we can verify that our a priori expectation 
to detect more anomalies of the second type when the merger trees become quite complex (i.e. trees of massive $z=0$ haloes composed of
many branches) is borne out. For the largest DPM haloes, this anomaly is bound to appear at least once:
the percentage of merger trees containing an anomaly of the second kind is close to 0 for the smallest haloes, 
and reaches 100 \% for the largest ones. This rise is quite slow until 4.8 $10^{14}$ M$_{\odot}$ where the 10 \% mark is reached, but 
quickly accelerates until, from 5.9 $10^{14}$ M$_{\odot}$ onward, all DPM trees contain the second kind of anomaly at least once. 
With the MSM, the second type of anomaly hardly ever occurs, meaning that the percentage of trees containing this anomaly is always close to 0, except for the 
4.5 $10^{14}$ M$_{\odot}$ mass bin where 1 tree out of 6 contains this anomaly. For the BHM, until 9 $10^{13}$ M$_{\odot}$, the percentage of trees containing this anomaly is below 5 \%. 
Interestingly enough, the number of second type anomalies for the BHM is higher than for the MSM. Around 4.5 $10^{14}$ M$_{\odot}$, one third of the trees are plagued with this
anomaly and above 6 $10^{14}$ M$_{\odot}$ this fraction rises to one half. This seemingly dramatic difference must be somewhat tempered by 
the small number of events as the two final mass bins contain eight trees in total. However, we conjecture that this behaviour of the BHM is 
the result of the propagation in time of a small fraction of the much more frequent anomalies of the third kind. To be more specific,
it so happens that the in-built tendency of the BHM to preserve the level of subhaloes from one output to the next leads, in some cases, to confuse subhalo
and halo when the branches finally merge together.   

As a matter of fact, looking at the occurrence of anomalies of the third kind displayed in Fig. \ref{trees_err3}, we notice that they appear much more frequently than the anomalies
 of the second kind (error bars at a given percentage are much smaller than in Fig. \ref{trees_err2}). Here again, as expected,
these anomalies become more frequent as merger trees become more complex. For the largest haloes, this anomaly is bound to appear at least once, whatever 
the method used to build the merger tree. For the DPM this rise is quite steady, the 50 \% threshold being reached for trees with final haloes of mass $\sim$ 7 $10^{12}$ M$_{\odot}$,
and from 9 $10^{13}$ M$_{\odot}$ onward, all trees contain this anomaly at least once. For the MSM method, the transition is more
pronounced, with the fraction of trees containing the anomaly rising quite slowly until $10^{13}$ M$_{\odot}$,  
reaching the 50 \% threshold for 2.5 $10^{13}$ M$_{\odot}$ haloes, and 100 \% at 1.7 $10^{14}$ M$_{\odot}$. 
However, the discrepancy between MSM and BHM is in favour of the latter this time around: the fraction of trees containing anomalies of the third kind gently rises from 
0 to 20 \% when final haloes masses attain 4.8 $10^{13}$ M$_{\odot}$, reaches 50 \% at 1.7 $10^{14}$ M$_{\odot}$ and 100 \%
at 3.2 $10^{14}$ M$_{\odot}$. In other words, the number of trees containing this anomaly is reduced by a factor 4 on average when going from the MSM to the BHM, 
so that the conversion of a few of these anomalies into anomalies of the second kind seems a small price to pay. Also,
 we have good reasons to believe that an extended
BHM over more than two time outputs could definitively resolve the issue.  

The conclusions we can draw from these three anomaly tests are that (i) the method used when creating the halo tree 
has an important impact on the merger tree (ii) the most obvious anomalies can be avoided by using an MSM like method 
(iii) further improvement is possible using BHM like methods but it does not reduce all anomalies to the same extent 
and comes at quite an expensive cost in terms of complexity of algorithm and CPU requirement. 
In the remaining of the paper we examine how the shape of individual merger trees is affected when one 
uses these three different tree building methods, illustrating how these differences can affect SAMs 
which are grafted on the trees.

\subsection{Examples of individual merger trees and their consequences in terms of halo formation epoch}

\begin{figure}[bp]
  \centering
  \includegraphics[width=9 cm]{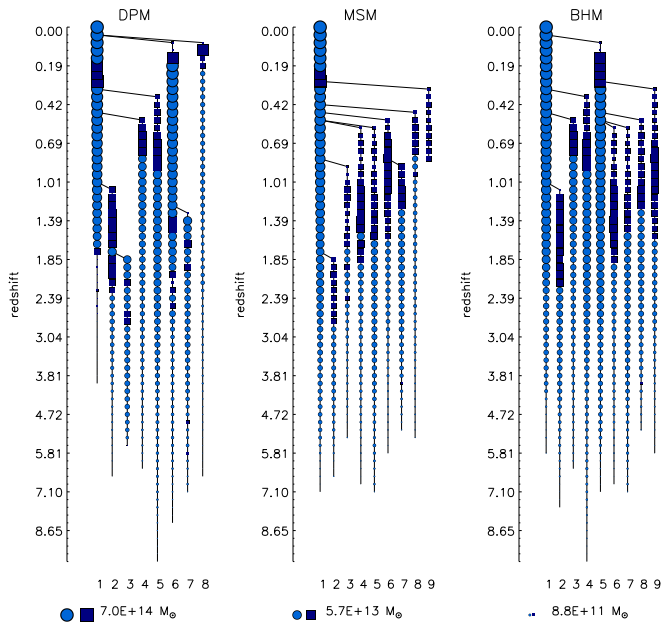}
  \caption{Zoom on merger trees obtained for the same final halo using three methods of subhalo selection. 
The DPM tree is shown on the left hand side, the MSM tree in the middle and the BHM tree on the right hand side 
(see text, Sects 3.1 and 3.3, for detail). Circles represent haloes, squares subhaloes. Solid lines show progenitor--son relationships, 
dashed lines stand for mergers that resulted in one of the merging haloes surviving as a subhalo.}
  \label{tree_zoom_18}
\end{figure}

\begin{figure}[ht]
  \centering
  \includegraphics[width=9 cm]{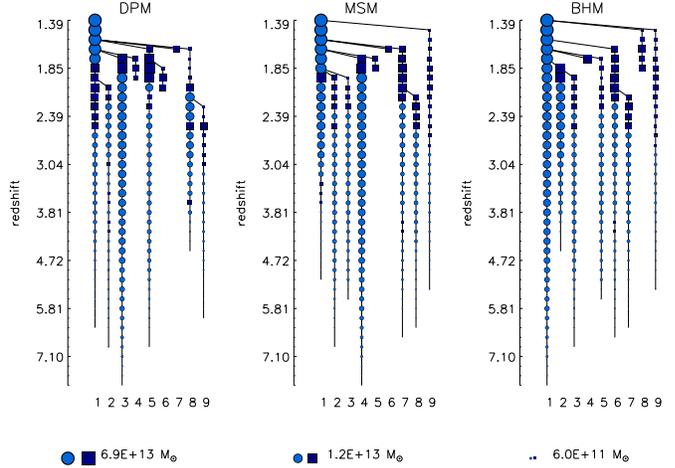}
  \caption{Same as Fig. \ref{tree_zoom_18}, for another halo.}
  \label{tree_zoom_23663}
\end{figure}

\begin{figure}[bp]
  \centering
  \includegraphics[width=9 cm]{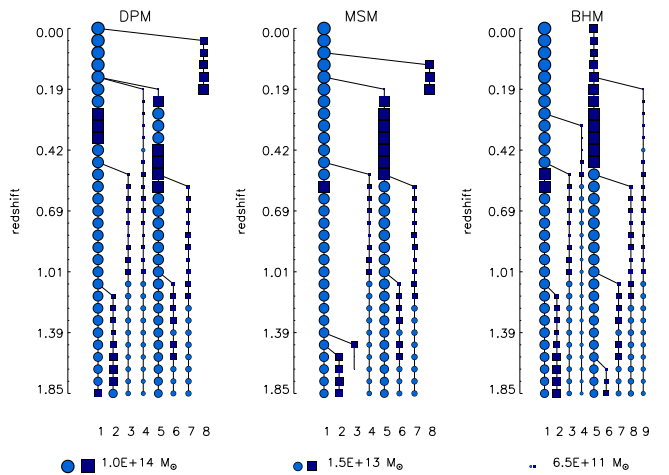}
  \caption{Same as Figs. \ref{tree_zoom_18} and \ref{tree_zoom_23663}, for a different halo.}
  \label{tree_zoom_21443}
\end{figure}

We pick merger trees where the three methods DPM, MSM and BHM differ and zoom in on the 
portion of the tree where these differences take place.
 For each Figs. \ref{tree_zoom_18}, \ref{tree_zoom_23663}, \ref{tree_zoom_21443} the left hand side 
displays the tree obtained using the DPM, the middle spot is occupied by the MSM tree and the BHM tree is 
shown on the right hand side. As in the previous plots throughout the paper, haloes are represented by circles and subhaloes 
by squares, solid lines show relationships between main progenitor and main son, dashed lines stand for mergers
where a halo survives as a subhalo and they link the main subhalo progenitor to its new host. For an easier analysis,
 each branch has been indexed from 1 to 9 (number below each branch).    

The first example (Fig. \ref{tree_zoom_18}) illustrates the dreadful effect the DPM can have on a merger tree. 
In the first (main) branch of the DPM tree, we can see an inversion between halo and subhalo (anomaly of the third kind) 
between branches 1 and 6 around redshift 0.15. Branch 1 also starts awkwardly as a small halo around redshift 3.9, keeps
its mass for about seventeen outputs, and suddenly becomes a much larger subhalo around $z=1.7$ without undergoing an obvious merger. 
This subhalo should be the remnant of the merger of branch 1 with branch 2, but instead an anomaly of the third kind appears.

The MSM tree for the same halo (middle tree in Fig. \ref{tree_zoom_18}) behaves in a much more civilised manner: the first (main) 
branch starts as expected with haloes growing in size regularly from one time output to the next. However, we still find a subhalo 
in the main branch around redshift 0.2. Looking back at the DPM tree, we can see that this is due to an identification problem 
with a branch that ends as a subhalo at redshift 0 in the MSM tree but not in the DPM tree (equivalent of DPM branch 6), and for 
this reason is not shown here (we only plot the merger tree of the main halo for each method). We notice that branch 3 has two
merger events with branch 5, leading to its main halo turning into a subhalo for one output on both occasions. 
Branch 6 merges with branch 7: it is a subhalo-subhalo merger where the resulting subhalo eventually merges with the main branch
around redshift 0.5. 
We can also see that branch 9 starts as a quite large subhalo around redshift 0.8, and thus can be defined as an anomaly of the first kind.
Nevertheless, we still conclude that for this halo, the MSM method is better than the DPM as the most obvious 
anomalies present in the DPM main branch are greatly suppressed by the MSM and 
the time evolution of subhaloes seems to be more accurate overall as well.
Moving to the right hand side of the figure, which shows the tree obtained with the BHM, we first notice that the
subhalo in the main branch has disappeared. Further comparing the BHM main branch to that of the MSM, 
we clearly see that (as was already the case for DPM and MSM), these differ from the origin onward, with the MSM branch 
appearing at $z=7$ and the BHM one later at $z=5.8$. The main branch of the MSM tree has become branch 5 of the BHM 
tree, and branches 4, 5, 6 and 7 of the MSM tree correspond to branches 6, 7, 9 and 8 of the BHM tree respectively. 
The BHM main branch is not seen in any of the other two main halo trees (it is a branch of one of their
subhaloes identified at $z=0$ whose trees are not represented on this plot) but 
we recognise that branches 3 and 7 in the BHM tree are branches 4 and 7 in the DPM tree.

The second example proposed in Fig. \ref{tree_zoom_23663} is also a zoom of a main branch but only from redshift 8 
down to redshift 1.4 this time. Once again subhaloes are present in the main branch of the DPM
 tree as a result of a merger with branch 3 around $z=2.4$.  
The subhalo in the MSM main branch at redshift 2, however is due to a type 3 anomaly involving branch 4. This latter is also branch 3 of the DPM tree and
the main branch of the BHM tree, which explains why the anomaly disappears in that case. 
The main branch of the MSM tree can be partly identified with branch 2 of the BHM tree, even though their first 
haloes (from $z=5$ to $z=3.2$) differ. 
Apart from this, MSM and BHM trees are very similar overall: we can identify branches 2, 7 ,8 and 9 of the MSM tree with branches 3, 6,7 and 9 of the BHM tree.

The last example (Fig. \ref{tree_zoom_21443}) illustrates that the BHM is not entirely fool proof, in the sense that not 
all anomalies in its merger trees can be eradicated. This example is a zoom on the main branch of a halo between redshifts $1.85$ 
and $0$. As we can see in the figure, several subhaloes appear in the main branch of the DPM tree. Their presence is caused by an 
exchange between haloes and subhaloes first with branch 2 ($z=1.85$) then with branch 5 ($z=0.42$). The second occurrence of 
this anomaly lasts 3 steps. 
In the MSM tree only one subhalo appears in the main branch. 
Except for branch 4 that does not appear and branch 3 which is branch 4 in the MSM tree, all the branches have the same index in both DPM and MSM trees. 

The first occurrence of the anomaly (between DPM branches 1 and 2) is prevented by going to the MSM method, however one 
occurrence of an anomaly of the third kind persists between branches 1 and 5. In both cases there is an anomaly of the first kind in branch 8. 
As these branches broadly have the same thickness, the merger between their haloes is a major one, which explains why the MSM
fails. Looking at the main branch of the BHM tree we see that it contains two subhaloes, which is surprising since this result
is worse than that obtained with the MSM method. It is, in fact not such an unexpected turn of events: 
the BHM simply performed as well as the MSM in the first instance when the subhalo appeared due to the major merger between the 
two haloes. However at the following time output, it took into account the information that the halo of the main branch had 
become a subhalo and since it was possible, decided to maintain its subhalo status.
At the following output, this possibility had vanished, and the subhalo was restored to its main halo status. 
Another difference between the MSM and the BHM trees is quite noticeable: branch 5 of the BHM tree corresponds 
to both MSM branches 5 and 8. The BHM method managed to preserve the progenitor--son link of the subhalo involved in the 
major merger two time outputs further than the MSM. 

\begin{figure}[tp]
  \centering
  \includegraphics[width=9 cm]{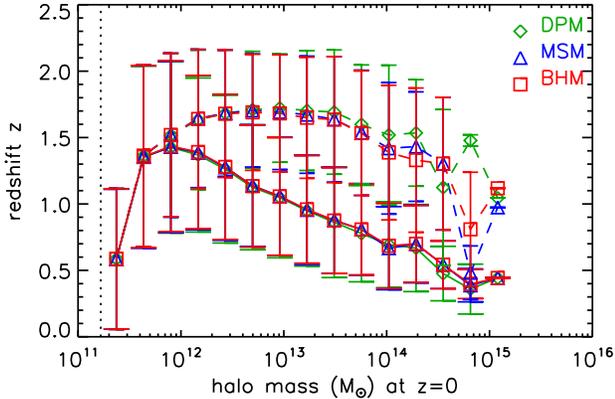}
  \caption{Formation and assembly epoch of $z=0$ main haloes according to the different merging tree building methods 
 described in the text (Sects. 3.1 and 3.3). Dashed (upper) curve shows formation redshift (defined as the redshift 
when the sum over the mass of all progenitors at a given time output reaches 50\% of the final 
$z=0$ main halo mass) as a function of halo mass, and solid (lower) curve shows assembly redshift 
(defined as the redshift where 50\% of the mass of the final main halo is assembled 
in the main branch for the first time).}
  \label{formation_time}
\end{figure}

Turning to a statistical measure to quantify the impact of using different methods to build halo merger trees, we now focus on 
measuring the ``downsizing/upsizing'' nature of the formation process of dark matter haloes. 
This is an important issue for galaxy formation as observations reveal (e.g. \citet{Thomas2005}) that most massive galaxies, 
which are generally located in the most massive DM haloes, are composed of older stars than their less massive 
counterparts. Our results are plotted in Fig. \ref{formation_time} for the 3 different methods we use to build 
merging trees. There are 2 sets of curves in this figure, the first one (lower curves) showing what we call the ``redshift 
of assembly'' ($z_a$) as a function of halo mass, and the second one (upper curves) showing the ``redshift of formation''
($z_f$). $z_a$ is defined as the redshift when 50\% of the mass of the $z=0$ main halo is assembled 
in the main branch (the branch of the main progenitor or trunk) for the first time. $z_f$ 
is the redshift at which the sum of the masses of {\em all} progenitors (independent of the branch they are part of) 
first reach 50\% of the mass of the $z=0$ main halo\footnote{If we use AdaptaHOP haloes instead of main haloes, i.e. 
include all substructure inside the main halo, the 2 curves we obtain are almost identical to these shown here 
for the main (substructureless) haloes, albeit the error bars are smaller in the higher mass bins for the redshift of formation.}.

The first thing to note is that all three methods yield very similar results with the exception of the formation 
redshift in the one before last mass bin, where the different identification of a single main halo by the methods is blown
 out of proportion by poor statistics. It is interesting to compare Fig. \ref{formation_time} 
to Fig. 5 of \citet{Neistein2006}, who plot the same quantities for Monte-Carlo merging trees based on the 
EPS formalism, as we can use their calculations to infer the impact of mass resolution on our results. 
As these authors point out, for haloes of a few 10$^{11}$ M$_\odot$ at $z=0$, i.e. close to our resolution limit, 
the number of progenitors is small, and the full merger tree is not much more complex than the main branch. 
This has two effects we can see: the values of $z_a$ and $z_f$ are (i) not very different from 
one another (ii) artificially drop at low masses. When the mass gets higher, the 
scatter in both $z_a$ and $z_f$ slowly decreases, even though it is not so clear from Fig. \ref{formation_time}
because using the main halo instead of the AdaptaHOP halo naturally increases the dispersion.    
Finally, whereas $z_a$ is very similar, $z_f$ in Fig. \ref{formation_time} is much lower than the 
average $z_f$ value plotted by \citet{Neistein2006} in their Fig. 5, peaking at $z_f=1.7$ instead of 
$z_f=5$. This possibly reflects the fact that our 
mass resolution is lower than theirs by two orders of magnitude at 10$^{11}$ M$_\odot$ compared to 10$^9$ M$\odot$. If so, it
is very instructive in terms of the impact of mass resolution on downsizing: if the epoch of half-mass formation
is such a sensitive function of mass resolution as it seems to be, it means that one could easily be a few billion years off
when estimating the mean age of stellar population of galaxies from a poorly resolved $N-$body simulation.
We note however that the theoretical estimate these authors plot in the same figure is flatter and peaks at a lower 
redshift of $z_f=3.5$, in much better agreement with our $N-$body simulation results.
Obviously this is less of a problem for $z_a$ since this measures the moment when 50\% of the mass is assembled 
in a unique object which is well resolved as soon as haloes are massive enough. 

In light of all the tests we have performed, we conclude that the MSM purges merging trees of many of the anomalies which 
take place when building them with the DPM. This simply confirms that when choosing
 a subnode to create a halo, one is best advised to use the most massive one instead of the densest. 
The BHM itself generally yields similar results as the MSM, however the extra information it gleans from the  
previous time output helps to get rid of most anomalies of the third kind, even though it
 does propagate an anomaly of the same kind in the tree from time to time. Statistical quantities 
like the average formation time and the mean redshift of assembly of dark matter haloes are fairly insensitive to the
method used to build the trees, but this is less true of the dispersion around this value. 

%
%______________________________________________________________
\section{Discussion and Conclusions}

We have presented a comprehensive study of the problems one necessarily encounters 
when building dark matter halo and subhalo merger trees directly from $N$-body simulations. 
We have also suggested methods to greatly reduce them, if not solve them perfectly.
To the best of our knowledge, the first trees of this kind were used by
\citet{Springel2005,De_LuciaBlaizot2007} to populate the millennium cosmological simulation with galaxies.
However, the issues we have described in this paper are not the main focus of any 
of these papers, and we strongly believe they needed to be addressed in detail, especially
since a lot of recent work has been devoted to comparisons between Monte-Carlo merger 
trees to the millennium trees \citep{NeisteinDekel2008a,Parkinson2008,Forero2009b}.

The first part of the paper was focused on the difference between dark matter (sub)haloes and their properties
(mass and number of objects) identified using AdaptaHOP and 
simple criteria to separate subhaloes from haloes (DPM, MSM).  
We also demonstrated that whereas the halo 
mass function is robust to a change in subhalo identification criteria,
there exists a relatively important effect of the latter on the estimate of the mass fraction contained 
in subhaloes. We concluded this study by arguing that, in keeping with the philosophy of using information contained in particle 
positions only to define haloes and subhaloes, one should use a mass criterion over a density 
one to split the internal structure of haloes apart.

This ensures not only that subhaloes identified in this way are 
always less massive than their host, but also that the time evolution of the subhalo (mass loss, encounters
with multiple fly-bys) can be better understood. Indeed using subhaloes in merger history trees 
helps to resolve issues such as halo fragmentation. Massive haloes which would have appeared
to have no progenitors at the previous time output can now be traced as well identified 
subhaloes in most cases. We also pointed out that using subhaloes in the merger history of haloes, apart from the 
obvious trend to make the merger history trees more complex because they have more branches,
have an impact on the merger histories of main haloes, in the sense that different methods to identify
subhaloes will yield different merging trees for the main halo as well. Finally, we demonstrated that 
as the physical meaning of a (sub)halo is linked to the merger history itself, using part of the past merger 
history of a (sub)halo to construct its future yields the best results of all methods used to build
merging history trees.

As far as mass resolution is concerned, while we did not make use of the most resolved simulations 
available to us, we feel that we were justified in doing so, as the problems we identified will simply 
be exacerbated when moving to higher resolution simulations. An interesting side issue that mass 
resolution raises is if/how much merging trees
are preserved when it increases. It was beyond the scope of this first paper, but will be the 
main focus of our next one. We also could have presented results where we varied time resolution (i.e. the number 
of outputs of the $N$-body simulation one uses to build the trees) but whilst 
occurrences of anomalies in the resulting merger trees would have increased with decreasing time 
resolution, our conclusions as to the best method to be employed to build the trees would stand.

Although we did not comment much about this point in the main body of the text, our work presumably has 
non-negligible implications on semi-analytic models of galaxy formation and evolution. More specifically,
the anomalies which occur in merger trees built using the different methods (DPM, MSM, BHM) will interact 
with recipes employed to estimate the cooling rate of hot gas and feedback from active galactic nuclei
in these models. Whilst this may not be important statistically for the population of galaxies 
considered as a whole, calculations of properties of specific sub-populations may be sensitive
to the issue. In particular, galaxies hosted by massive haloes will very likely suffer as these 
are the places where these anomalies are most common: we plan to address this question in a future 
work.

Finally, the main advantage to be gained using trees directly constructed from $N$-body simulations 
to graft on SAM of galaxy formation and evolution, is clustering information. Monte-Carlo 
merging trees not only do not contain position information but they also fail
to describe environment effects on low mass halo assembly \citep{ShethTormen2004,Wechler2006,Reed2007,Zhu2006}. 
Positioning SAM galaxies within subhaloes allows one  
to both compute merging timescales more accurately, but also to measure the intra halo 
correlation functions predicted by the model with great accuracy \citep{Blaizot2006}.
These previous statements need to be somewhat toned down since baryons exert a gravitational effect on 
dark matter and this will very likely impact the survival of subhaloes, however these 
effects can be quantified using the framework we have developed in this paper and we plan to address
the issue in the near future.

\bibliography{bibliography}

\begin{acknowledgements}
  DT acknowledges the support received from the European Science
  Foundation (ESF) for the activity entitled 'Computational Astrophysics and
  Cosmology' (Grant ASTROSIM ref 1974) during his stay at Oxford.
  We thank C. Rimes, C. Pichon and J. Forero-Romero for useful discussions.
  We also thank the referee for his corrections and comments.
  Simulations presented in this paper were run at the IDRIS.
  This work is part of the Horizon project.\\
  http://www.projet-horizon.fr/
  
\end{acknowledgements}
%
%______________________________________________________________

\newpage

\begin{appendix}
  
\section{Comparing halo finders: AdaptaHOP and FOF}

We first proceed in section A.1 of this appendix to compare AdaptaHOP haloes as defined at
the end the section 2.1 to
haloes identified using the popular FOF algorithm with a standard value of $b=0.2$
for the linking length parameter. To be consistent, we also choose
20 as the minimum number of particles a halo detected by the FOF algorithm can contain. 
The halo finder algorithms are then run on the same $N$-body simulation 
described in paper GalICS I \citep{Hatton2003}. We also find necessary to
compare the haloes and subhaloes obtained with AdaptaHOP using the MSM method
to the halos detected using the FOF algorithm, this task is summarised in section A.2.

\subsection{Comparing halos}

\subsubsection{Individual examples}

Our first goal is to ``calibrate'' AdaptaHOP. To do so, we simply collapse the node 
structure tree onto its first node to 
define a single halo as a group of particles with a density above the $\rho_t$
threshold.

In this appendix we refer to objects detected with AdaptaHOP and 
whose hierarchy of inner local density maxima has been collapsed 
as described in section 2.1 as AdaptaHOP haloes. Similarly 
FOF haloes are objects detected using the FOF algorithm.

\begin{figure}[ht]
  \centering
  \includegraphics[width= 9 cm]{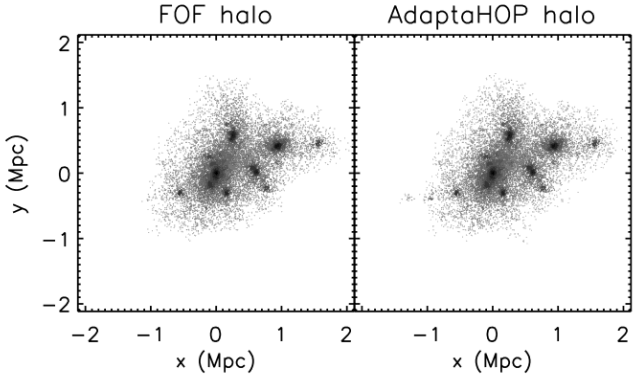}
  \caption{Example of a ``relaxed'' halo detected in our test simulation, the left panel
    is the halo detected using friend-of-friend (FOF) algorithm, the right
    panel is the same halo detected using AdaptaHOP. The left panel corresponds
    to the example shown in the top right panel of Fig.
    \ref{halo_MSM_48} before the subhalo decomposition of the AdaptaHOP halo.}
  \label{fof_adap_48}
\end{figure}

Intuitively, we expect haloes detected with FOF and AdaptaHOP to resemble one another, provided
they are fairly relaxed and well resolved, i.e. they are close to spherical in shape and contain a 
large enough number of particles. This is the case for the halo represented in 
Fig. \ref{fof_adap_48}. The projected positions of dark matter particles belonging to this halo 
in the (xy) plane are shown in both panels of this figure, with the centre of the halo located at point 
(0,0). On the left hand side panel we can see the FOF halo, on the right hand side panel, we can see 
the AdaptaHOP halo. At first glance, one can easily be convinced that they are mostly the same halo, 
but, looking a bit closer, one gets the impression that the AdaptaHOP 
halo contains a few small overdensities which are not included in the FOF halo.

\begin{figure}[ht]
  \centering
  \includegraphics[width= 9 cm]{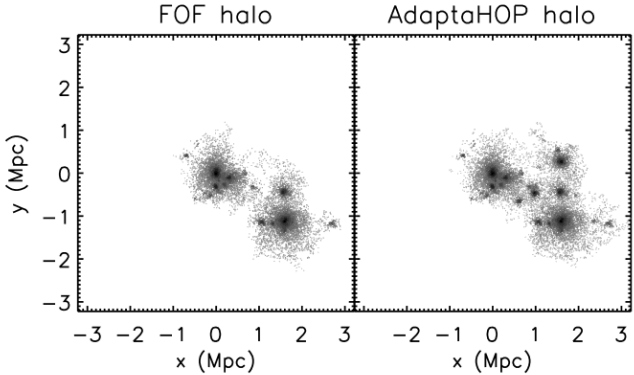}
  \caption{Example of a ``merging'' halo detected in our test simulation, the left panel
    is the halo detected using friend-of-friend (FOF) algorithm, the right
    panel is the same halo detected using AdaptaHOP. The left panel corresponds to the example shown in the top right panel of Fig.
    \ref{halo_MSM_1421} before the subhalo decomposition of the AdaptaHOP halo.}
  \label{fof_adap_1421}
\end{figure}

This impression is confirmed by going to 
the next halo example displayed in Fig. \ref{fof_adap_1421}. 
This one is ``peanut'' shaped as it has just undergone a merger.
Again both haloes are quite similar but the AdaptaHOP halo clearly shows 
overdensities which are not part of the FOF halo. 

From these two examples, we conclude that haloes detected by the two algorithms do not 
seem to be very sensitive to the dynamical state the halo is in. This is understandable since 
both methods rely on particle positions only to define haloes. 
However, the smoothing process used by algorithms which compute the density field
(AdaptaHOP in our case) seems responsible for the systematic inclusion of more overdensities
within the haloes than percolation algorithms (FOF here), especially when haloes 
are less relaxed. Even though this last point seems minor,
it is more important than it seems: if we wished to compare the
SUBFIND and AdaptaHOP subhalo distribution for instance, SUBFIND would not detect 
these extra AdaptaHOP overdensities since it performs a FOF first step, and this would, 
in turn, influence the computation of the potential energy of the particles that SUBFIND 
performs, possibly resulting in a different stripping of particles not bound to the halo.

\subsubsection{Comparing halo finders: statistics.}

The two individual halo examples discussed in the previous section
naturally lead us to wonder about how general the conclusions we have drawn 
really are. In other words:
\begin{itemize}
\item How well does the mass
  distribution of resolved haloes obtained with both halo finders agree? 
\item How frequent is the splitting of single AdaptaHOP haloes into several FOF haloes?
      Does it depend on resolution/mass? Are all pieces detected?
\item How do the smallest, poorly resolved, AdaptaHOP and FOF halo (which do not
show internal overdensities) populations compare?
\end{itemize}

\begin{figure}[ht]
  \centering
  \includegraphics[width=8 cm]{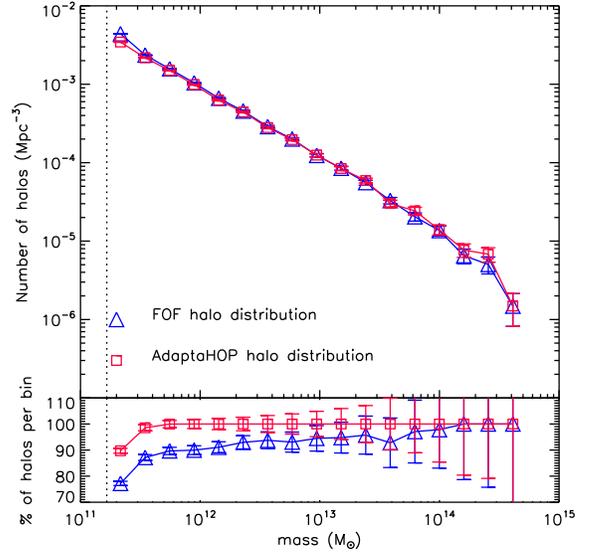}
  \caption{For both halo finders, haloes detected at redshift 0 were sorted into 25 mass bins. The
  top panel shows the number of haloes detected with FOF (triangles) and
  AdaptaHOP (squares) in each mass bin. 
  Using their particle content, we then cross-identify FOF haloes with their AdaptaHOP counterparts, 
  as described in the text (section 2.1.3). In other words, we enforce that each FOF halo is identified with at most one AdaptaHOP halo 
  and that the identification yields the same result when performed in the reverse order.
  The bottom panel displays the percentage of haloes per mass bin thus cross identified.
  The error bars correspond to Poisson uncertainties.
  The vertical dotted line present on both panels corresponds to the 20 particle detection threshold.}
  \label{mass_function_FOF_HOP}
\end{figure}

The mass distributions of haloes at redshift 0 are shown in Fig. \ref{mass_function_FOF_HOP}. In both
 panels squares correspond to the FOF halo distribution and
triangles to the AdaptaHOP halo distribution. The mass threshold of 20
particles is represented by a vertical dashed line. The first impression one gets from the top panel of the
figure is that the mass distributions of haloes are very close indeed. 
Looking a bit closer, we find that the number of FOF haloes is more than 5\% higher than the number 
of AdaptaHOP haloes in the first 5 bins, with about 15000 FOF haloes against 11600 AdaptaHOP haloes 
of a mass lower than 2.67 $10^{11}$ M$_{\odot}$. For masses above
4.5 $10^{11}$ M$_{\odot}$, both algorithms converge within 0.5\%, as we count 1842 FOF haloes and 1856 AdaptaHOP haloes.

To compare FOF haloes to their AdaptaHOP counterparts, we simply use the list of particles 
belonging to each halo as follows:
(i) an AdaptaHOP halo is identified with at most one FOF halo, the one which contains the highest 
fraction of its particles (ii) the same procedure is applied to identify a FOF halo to 
its one and only AdaptaHOP counterpart. A FOF halo and its AdaptaHOP counterpart are then deemed 
to be the same object when both 
the highest fraction of the AdaptaHOP halo particles are found in the FOF halo, and  the highest fraction of 
the FOF halo particles are found in the AdaptaHOP halo. For each mass bin we then compute the number of haloes  
which are found to be ``identical'' in that sense. The curves in the bottom panel 
are obtained by dividing this 
number by the number of FOF haloes in the bin (triangles) or by the number of 
AdaptaHOP haloes in the bin (square), for each mass bin.
 This yields the percentage of haloes from each algorithm (FOF or AdaptaHOP) also identified as single haloes by the other. In the case of the least resolved haloes, 90 \% of AdaptaHOP haloes are identified as FOF haloes, 
but this percentage drops to 77 \% for FOF haloes identified as AdaptaHOP haloes. 
Also in the former case, the percentage hits the 100 \% mark from 5 $10^{12}$ 
M$_{\odot}$ onward, whereas for the latter the percentage rises more slowly, reaching 90 \%  for haloes around 5 $10^{12}$ M$_{\odot}$, and 100 \% for haloes of $10^{14}$ M$_{\odot}$ only.

\begin{figure}[ht]
  \centering
  \includegraphics[width=8 cm]{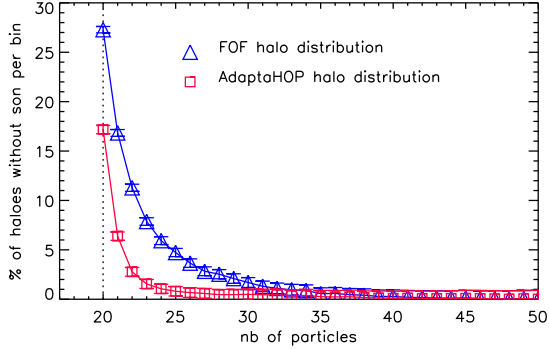}
  \caption{Percentage per bin (of width 1 particle) of haloes with less than 50 particles 
  which are not detected in two consecutive snapshots between redshifts 0 and 10, i.e. 
  among a total of 70 snapshots.
  The error bars correspond to Poisson uncertainties. 
  The vertical dotted line corresponds to the 20 particle detection threshold.}
  \label{small_halo_lost}
\end{figure}

Whilst one can argue that the comparison
of the two algorithms is irrelevant in a regime where results should not be trusted because of a too small number of particles
per halo, we believe it is nevertheless important for two reasons: (i) it better underlines the differences between the algorithms and (ii) semi-analytic models of galaxy formation do populate such low resolution haloes with galaxies (e.g. \citet{Hatton2003, Springel2005}).
The consequent discrepancy at the low mass end is quite worrisome,  and while it is well known that all halo
finders will be incomplete when the number of neighbours used to smooth the density field is close to the minimum number of particles per halo, one may wonder whether a significant fraction of these haloes is in any case marginally bound.

A good indicator of this (other than the measure 
of the total energy of the halo which has quite a large error bar attached to it) is their stability in time.
This data is shown in Fig. \ref{small_halo_lost}. We find that 27 \% of the
haloes containing 20 particles detected by FOF at a given time output are not found 
at the next. This number drops to 17 \% for AdaptaHOP haloes.
For both halo finders this fraction decreases quickly when the number of particles
per halo increases. In other words, when using the FOF, we need to reach a resolution of at least 
34 particles to lose less than 1 \% of the haloes between time outputs. In the case of AdaptaHOP, 
with a resolution of 24 particles or more, less than 1 \% of haloes are lost. From these numbers, 
we conjecture that only a small part of the discrepancy between the two algorithms at the low mass end of the halo mass 
function comes from the fact that AdaptaHOP (and in general algorithms using density criteria) is 
more efficient than FOF (i.e. algorithms using percolation criteria) for selecting bound objects.  

\begin{figure}[ht]
  \centering
  \includegraphics[width=8 cm]{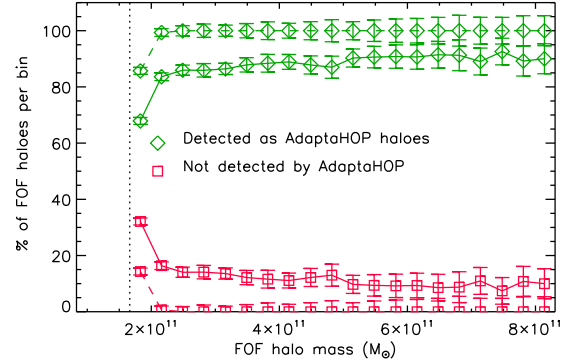}
  \caption{Every FOF halo detected at redshift 0, with less than 100 particles, 
is identified with its AdaptaHOP counterpart using the method described in the text (section 2.1.3). Solid curves show the case
where we enforce that a unique FOF halo (the one which contains the largest number of particles of 
its AdaptaHOP counterpart) be identified with its best AdaptaHOP counterpart (the one which contains 
the highest fraction of the FOF halo particles), as in Fig.
\ref{mass_function_FOF_HOP}. 
Dashed curves show what happens when we relax this constraint, and allow several FOF haloes to be 
identified with the same AdaptaHOP counterpart.
The error bars correspond to Poisson uncertainties.
    The vertical dotted line corresponds to the 20 particles detection threshold.}
  \label{detection_status100_HOP}
\end{figure}

\begin{figure}[ht]
  \centering
  \includegraphics[width=8 cm]{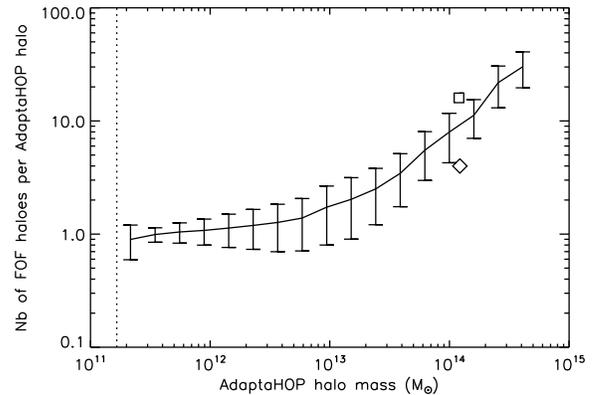}
  \caption{Average number of FOF haloes at redshift 0 per AdaptaHOP halo.
    The error bars correspond to the mean quadratic dispersion.
  The vertical dotted line corresponds to the 20 particles detection threshold. The diamond and the square 
  correspond to the 2 example haloes shown in Figs. \ref{fof_adap_48} and \ref{fof_adap_1421} respectively.}
  \label{nfof_adap}
\end{figure} 

To further substantiate this claim, we now proceed to check the other possible source of 
discrepancy, i.e. we address the issue of how many of the smallest FOF haloes
are not identified as individual haloes by AdaptaHOP, but simply detected as parts of other (larger) haloes.
We therefore apply the same technique we used to produce Fig. \ref{mass_function_FOF_HOP} but only 
to FOF haloes with less than 100 particles since we are interested in the low mass end, and 
relaxing the constraint that an AdaptaHOP halo must necessarily have a unique FOF counterpart. 

Results of this experiment 
are shown as dotted curves in Fig. \ref{detection_status100_HOP}. Examination of this figure 
reveals that:
\begin{itemize}
\item when the constraint of unicity is enforced (solid curve), 
the percentage of FOF haloes identified as individual 
AdaptaHOP haloes never reaches 100 \%, even for haloes containing 100 particles. Instead, this percentage 
rises slowly from  68 \% for FOF haloes in the first (20-24 particles) mass bin to reach 90 \% (+/- 1 \%) 
for FOF haloes more massive than 4 $10^{11}$ M$_{\odot}$  (44 particles or so).
\item when the constraint of unicity is relaxed (dashed curve) --- meaning that one AdaptaHOP halo can
 have several FOF counterparts ---, only 14 \% of FOF haloes in the first mass bin (20-24 particles) are not 
identified at all by AdaptaHOP as being part of a halo, and this percentage drops to less 
than 1 \% in the next mass bin (24-28 particles). Furthermore, from a FOF halo mass of 2.65 $10^{11}$ 
M$_{\odot}$ (32 particles) onward, not a single FOF halo remains undetected by AdaptaHOP.
\end{itemize}

These numbers lead us to conclude that most of the discrepancy of the halo mass function 
between halo finders at the low mass end can indeed be attributed to a different 
integration of small objects into larger ones, i.e. to different numbers of small 
overdensities being included in FOF haloes and their AdaptaHOP
counterparts. 

One cannot help to wonder how mass dependent such a statement is, i.e. if these small haloes preferentially 
get included in the outskirts of large clusters, or if their ``mis''-classification happens 
uniformly across the whole halo mass range spanned by the $N$-body simulation. 
This information is presented in Fig. \ref{nfof_adap}, where
we counted the number of FOF haloes detected at the last time output ($z=0$) inside
each halo found by AdaptaHOP. The AdaptaHOP halo population was then split into 25 mass bins, and for each 
of these bins we computed the average number (and  associated mean quadratic dispersion) of FOF haloes detected
as counterparts of the AdaptaHOP haloes. For smallish, galaxy size AdaptaHOP haloes (less massive than 
4 $10^{12}$  M$_{\odot}$ or about 400 particles), we find that the average number of FOF haloes per AdaptaHOP 
halo stays close to the minimal value of one, but starts 
increasing quite steeply as a function of AdaptaHOP halo mass after that, scaling as $N \propto M^{3/4}$. 
This implies that cluster size haloes detected by AdaptaHOP
with masses comprised between $10^{14}$ and  $10^{15}$ M$_{\odot}$ contain up to several tens of
FOF haloes. Since the halo mass function decays as $N \propto M^{-1}$ (see Fig. \ref{mass_function_FOF_HOP})
in this mass range, this means that cluster size AdaptaHOP haloes get in total roughly the same number of small 
``undetected'' FOF haloes as galaxy/group size AdaptaHOP haloes, only they are redistributed over a 
smaller number of objects. As an illustration of the scatter around the average value, the two example 
haloes of similar masses displayed in figures \ref{fof_adap_48} and \ref{fof_adap_1421} are marked 
with a diamond and a square in Fig. \ref{nfof_adap} respectively, with the diamond AdaptaHOP halo containing 
3 FOF haloes and the square one 16. Note that this explains why in those two cases, most, if not all the 
extra ``clumps'' of particles seen in the AdaptaHOP halo are not present in the FOF halo to which it was 
identified, but are detected by FOF as stand alone haloes.  

The two group finding algorithms we have studied (FOF and AdaptaHOP) give similar results. 
The discrepancies measured between them occur principally at the low mass end of the halo mass 
function and can be brought down to the percent level when one considers that haloes detected 
using AdaptaHOP generally contain several FOF haloes. Alternatively, one could use
(as in the HOP algorithm, see \citet{Eisenstein1998}), multiple/adaptive density thresholds to separate 
haloes from the background to reduce discrepancies with percolation algorithms. However it is unclear
that the various thresholds to use can be determined using physical arguments, so we believe 
it makes more sense to limit the number of ``free'' parameters and use a single density threshold, 
as implemented in AdaptaHOP. One can then take advantage
 of the embedded hierarchy of density maxima within each of the haloes identified that way to define subhaloes. 
Finally, we emphasise that it is hard to decide which of the algorithms (percolation or density based)
is closer to the ``true'' gravitationally bound halo, as poorly resolved substructures located in the 
outskirts of massive haloes are neither isolated, nor fully relaxed objects, and as 
such, a reliable estimate of their total energy is not easy to achieve, making it very 
difficult to decide whether they are bound to the larger halo.

\subsection{Comparing halo finder MSM vs FOF}

\subsubsection{Individual example}

\begin{figure}[ht] 
  \centering
  \includegraphics[width= 9 cm]{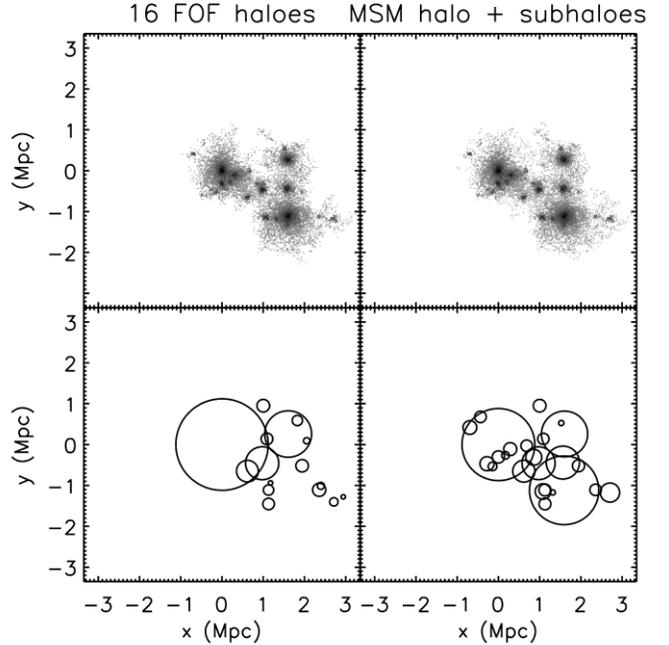}
  \caption{Top right panel: AdaptaHOP halo shown in the right panel of Fig. \ref{fof_adap_1421}. Top left panel:
 the 16 FOF haloes whose particles are found in this AdaptaHOP halo (the best FOF halo match is shown in the 
 left panel of Fig. \ref{fof_adap_1421}). Bottom left panel: virial regions of the 16 FOF haloes. 
 Bottom right panel: virial regions of the MSM halo and subhaloes as detailed
 in Fig. \ref{halo_MSM_1421}.}
  \label{cluster_fof_adp_1421}
\end{figure}

As mentioned in the previous subsection of this appendix, by mapping particles
contained in an AdaptaHOP halo onto the FOF halo distribution, one realises that
these particles often belong to several FOF haloes. This is illustrated in Fig. 
\ref{cluster_fof_adp_1421}, where we see in the top left panel the 16 FOF haloes which contain 
particles from the single AdaptaHOP halo plotted in the top right panel.
Recall that this means that for each of these 16 FOF haloes most of their particles 
are cross identified as particles belonging to the AdaptaHOP halo. 
Each circle in the bottom left
panel represents the virial sphere of one of these 16 FOF haloes. The circles on the
bottom right panel represent the virial sphere of the main halo and its subhaloes defined 
using the MSM method. We point out that the
largest FOF halo in the cluster (shown in Fig. \ref{fof_adap_1421}) includes both the main MSM halo and
its largest subhalo. It means that from the FOF point of view the major merger between 
halo and largest subhalo has occurred as well. We also see that the 15 extra FOF haloes
are detected as MSM subhaloes of the main halo. As a matter of fact this cluster of 16 FOF haloes 
is identified with a single structure tree containing 23 subhaloes by 
AdaptaHOP/MSM.

\subsubsection{Statistics}

Turning now to the issue of the identification of several FOF haloes with a single AdaptaHOP halo, we use the 
MSM method to check how often FOF haloes are detected as subhaloes rather than main haloes. 

\begin{figure}[ht]
  \centering
  \includegraphics[width=8 cm]{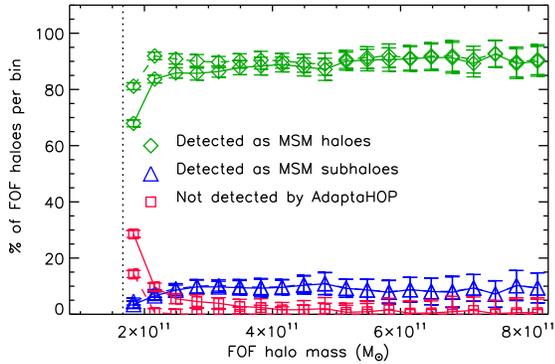}
  \caption{Cross identification of FOF haloes with less than 100 particles at $z=0$ and MSM haloes or subhaloes 
  (see text section 2.1.3 for details). In the first case (solid curves), only one FOF halo is allowed to be identified with 
  a MSM halo or subhalo. In the second case (dashed curves), several FOF haloes can be identified with a single 
   MSM halo or subhalo.
   The error bars correspond to Poisson uncertainties.
    The vertical dotted line corresponds to the 20 particles detection threshold.}
  \label{detection_status100_MSM}
\end{figure}

As we did in the section A.1.2 of this appendix, we now proceed to check that the conclusions 
we have just drawn for a single, well resolved halo are valid for the whole distribution 
of haloes, and especially for the poorly resolved ones. 
In order to do so, for each time output of the $N$-body simulation, we check whether 
each halo detected by the FOF is detected either as a halo or as a subhalo with the MSM.
This data is presented in Fig. \ref{detection_status100_MSM} for all haloes between  1.66 $10^{11}$
M$_{\odot}$ and 8.33 $10^{11}$ M$_{\odot}$ (i.e. 20 to 100 particles).
The solid curves in this figure are obtained when we allow only one FOF halo to be cross-identified with a MSM 
halo or subhalo. The percentage of FOF haloes identified as AdaptaHOP haloes (i.e. containing their 
subhaloes) is the same as in Fig. \ref{detection_status100_HOP}, but the main difference is that now the
smallest FOF haloes can be identified as subhaloes of larger haloes instead of having to be cross-identified
with a single small AdaptaHOP halo.
So the new information that we get from this figure, as compared to Fig. \ref{detection_status100_HOP} is   
that the percentage of FOF haloes identified as MSM subhaloes (triangles) is
quite small (5 \%) at the 20 particles mass threshold, increases
until 3 $10^{11}$ M$_{\odot}$ and remains steady around the 10 \% level up to 8 $10^{11}$ M$_{\odot}$. 
When we allow for the fact that several FOF haloes can be found in one MSM halo or subhalo we obtain the dashed 
curves, i.e. if one MSM halo or subhalo was cross-identified with  
at most one FOF halo the dashed and solid curves would perfectly match. 
The interesting result here is that this actually happens to be the case for all but the lowest FOF halo 
mass bins for MSM subhaloes (triangles), but not for MSM main haloes (diamonds) which means that small FOF
haloes are preferentially fused into the smooth component of the main halo found by the MSM method rather 
than with MSM subhaloes. 

\begin{figure}[ht]
  \centering
  \includegraphics[width=8 cm]{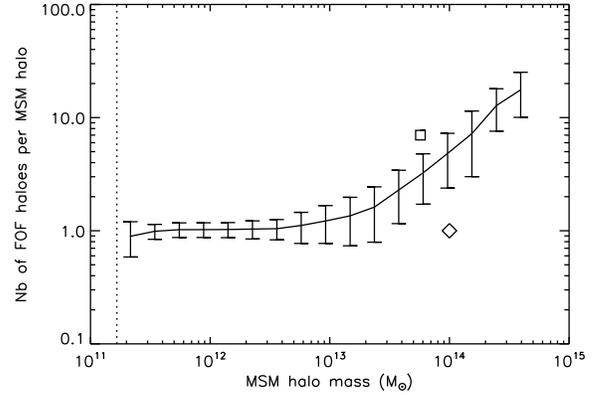}
  \caption{Number of FOF haloes at redshift 0 whose particles are found in a single
  main MSM halo (i.e. in an AdaptaHOP halo without its MSM subhaloes).
  The error bars correspond to the mean quadratic dispersion.
  The vertical dotted line corresponds to the 20 particles detection threshold. 
  The diamond and the square correspond to the examples shown in figures \ref{fof_adap_48} and 
  \ref{fof_adap_1421} respectively.}
  \label{nfof_MSM}
\end{figure} 

To quantify this behaviour a bit further,
we plot in Fig. \ref{nfof_MSM}, the average number of FOF haloes per MSM main halo. This figure was obtained in 
the same way as Fig. \ref{nfof_adap} except that we excluded FOF haloes cross-identified with MSM subhaloes. 
We notice that the average number of FOF haloes per MSM halo is close to 1 until we consider MSM haloes 
with masses greater than $10^{13}$ M$_{\odot}$. Then the average number of FOF haloes per MSM halo rises quickly 
as a function of MSM halo mass. At 4 $10^{14}$ M$_{\odot}$ we obtain on average 17.6 FOF haloes per MSM haloes. 
This means that small FOF haloes (less than 50 particles or so) are preferentially blended with the smooth main
halo component of group to cluster size main MSM haloes: they are not dense enough to be detected
as separate local maxima which is the necessary condition to be identified as subhaloes.

\end{appendix}

\end{document}